\documentclass[12pt,preprint,tighten]{aastex}

\def\etal{et al{.}}

\shorttitle{Type Ia SNe at $z\sim 0.5$}
\shortauthors{Clocchiatti et al.}

\begin{document}

\title{Hubble Space Telescope and Ground-Based Observations of 
Type Ia Supernovae at Redshift 0.5: Cosmological 
Implications\altaffilmark{1,2}}

\author{
Alejandro Clocchiatti,\altaffilmark{3} 
Brian P. Schmidt,\altaffilmark{4}
Alexei V. Filippenko,\altaffilmark{5}
Peter Challis,\altaffilmark{6}
Alison L. Coil,\altaffilmark{5}
R. Covarrubias,\altaffilmark{7}
Alan Diercks,\altaffilmark{7}
Peter Garnavich,\altaffilmark{8}
Lisa Germany,\altaffilmark{4}
Ron Gilliland,\altaffilmark{9}
Craig Hogan,\altaffilmark{7}
Saurabh Jha,\altaffilmark{6,5}
Robert P. Kirshner,\altaffilmark{6}
Bruno Leibundgut,\altaffilmark{10}
Doug Leonard,\altaffilmark{5,11}
Weidong Li,\altaffilmark{5}
Thomas Matheson,\altaffilmark{6,12}
Mark M. Phillips,\altaffilmark{13}
Jos\'e Luis Prieto,\altaffilmark{14,15}
David Reiss,\altaffilmark{7}
Adam G. Riess,\altaffilmark{9}
Robert Schommer,\altaffilmark{14,16}
R. Chris Smith,\altaffilmark{14}
Alicia Soderberg,\altaffilmark{5,11}
Jason Spyromilio,\altaffilmark{10}
Christopher Stubbs,\altaffilmark{17}
Nicholas B. Suntzeff,\altaffilmark{14}
John L. Tonry,\altaffilmark{18}
and
Patrick Woudt\altaffilmark{10,19}\\
\center{Accepted for publication in Astrophysical Journal}
}

\altaffiltext{1}{Based in part on observations with the NASA/ESA \emph{Hubble
Space Telescope,} obtained at the Space Telescope Science Institute,
which is operated by the Association of Universities for Research in
Astronomy (AURA), Inc., under NASA contract NAS 5-26555. This
research is primarily associated with proposal GO-7505.}
\altaffiltext{2}{
Based in part on observations taken at the Cerro Tololo Inter-American
Observatory.
Some of the data presented herein were obtained at the W. M. Keck
Observatory, which is operated as a scientific partnership among
the California Institute of Technology, the University of California,
and the National Aeronautics and Space Administration.  The Observatory
was made possible by the generous financial support of the W. M. Keck
Foundation.
Based in part on observations with the University of Hawaii 2.2-m telescope
at Mauna Kea Observatory, Institute for Astronomy, University of
Hawaii.
Based in part on observations obtained at the European Southern
Observatory, Paranal and La Silla, Chile, under programs
ESO 64.O-0391 and ESO 64.O-0404.
Based in part on observations taken at the WIYN Observatory, a
joint facility
of the University of Wisconsin (Madison), Indiana University, Yale
University, and the National Optical Astronomy Observatories.}
\altaffiltext{3}{Pontificia Universidad Cat\'{o}lica de Chile,
Departamento de Astronom\'{\i}a y Astrof\'{\i}sica,
Casilla 306, Santiago 22, Chile; {aclocchi@astro.puc.cl}.}
\altaffiltext{4}{The Research School of Astronomy and Astrophysics,
The Australian National University, Mount Stromlo and Siding Spring
Observatories, via Cotter Rd, Weston Creek PO 2611, Australia;
{brian@mso.anu.edu.au}.}
\altaffiltext{5}{Department of Astronomy, 601 Campbell Hall, University of
California, Berkeley, CA 94720-3411; {alex@astro.berkeley.edu}, 
{acoil@astro.berkeley.edu}, {weidong@astro.berkeley.edu}, {sjha@astro.berkeley.edu}.}
\altaffiltext{6}{Harvard-Smithsonian Center for Astrophysics, 60 Garden 
Street, Cambridge, MA 02138; {kirshner@cfa.harvard.edu}, {pchallis@cfa.harvard.edu}.}
\altaffiltext{7}{University of Washington, Department of Astronomy, Box 351580, 
Seattle, WA 98195-1580; {hogan@astro.washington.edu}.}
\altaffiltext{8}{University of Notre Dame, Department of Physics, 225 Niewland Science
Hall, Notre Dame, IN 46556-5670; {pgarnavi@nd.edu}.}
\altaffiltext{9}{Space Telescope Science Institute, 3700 San Martin Drive,
Baltimore, MD 21218; {rgillila@stsci.edu}, {ariess@stsci.edu}.}
\altaffiltext{10}{European Southern Observatory, Karl-Schwarzschild-Strasse 2, Garching, 
D-85748, Germany; {bleibund@eso.org}, {jspyromi@eso.org}.} 
\altaffiltext{11}{Department of Astronomy, MS 105-24, California Institute of Technology,
Pasadena, CA  91125; {leonard@astro.caltech.edu}.}
\altaffiltext{12}{National Optical Astronomy Observatory, 950 N. Cherry Ave.,
  Tucson, AZ 85719; {matheson@noao.edu}.}
\altaffiltext{13}{Las Campanas Observatory, Casilla 601, La Serena, Chile; {mmp@lco.cl}.}
\altaffiltext{14}{Cerro Tololo Inter-American Observatory, Casilla 603, La Serena, Chile;
{csmith@ctio.noao.edu}, {nsuntzeff@noao.edu}.}
\altaffiltext{15}{Ohio State University, Department of Astronomy,
4055 McPherson Laboratory, 140 W. 18th Ave., Columbus, Ohio 43210;
{prieto@astronomy.ohio-state.edu}.}
\altaffiltext{16}{Deceased 12 December 2001.}
\altaffiltext{17}{Department of Physics and Department of Astronomy,
17 Oxford Street, Harvard University, Cambridge MA 02138; {cstubbs@fas.harvard.edu}.}
\altaffiltext{18}{Institute for Astronomy, University of Hawaii, 2680 Woodlawn 
Drive, Honolulu, HI 96822; {jt@ifa.hawaii.edu}.
\altaffiltext{19}{Department of Astronomy, University of Cape Town, Private Bag,
Rondebosch 7700, South Africa, {pwoudt@circinus.ast.uct.ac.za}.}
}
\begin{abstract}

We present observations of the Type Ia supernovae (SNe) 1999M, 1999N, 1999Q, 1999S,
and 1999U, at redshift $z \approx 0.5$.
They were discovered in early 1999 with the 4.0~m Blanco telescope
at Cerro Tololo Inter-American Observatory by the High-$z$ Supernova
Search Team (HZT) and subsequently followed with many ground-based telescopes.
SNe 1999Q and 1999U were also observed with the
{\it Hubble Space Telescope}.
We computed luminosity distances to the new SNe using two methods,
and added them to the high-$z$ Hubble diagram that the HZT has been constructing
since 1995.

The new distance moduli confirm the results of previous work.
At $z \approx 0.5$, luminosity distances are larger than those
expected for an empty universe, implying that a
``Cosmological Constant,'' or another form of ``dark energy,'' has been
increasing the expansion rate of the Universe during the last few
billion years.

Combining these new HZT SNe~Ia with our previous results
and assuming a $\Lambda$CDM cosmology, we estimate the
cosmological parameters that best fit our measurements.
For a sample of 75 low-redshift and 47 high-redshift SNe~Ia with MLCS2k2
(Jha et al. 2005) luminosity calibration we obtain
$\Omega_M=0.79^{+0.15}_{-0.18}$ and
$\Omega_\Lambda= 1.57^{+0.24}_{-0.32}$ ($1\sigma$ uncertainties)
if no constraints are imposed, or
$\Omega_M = 0.29^{+0.06}_{-0.05}$
if $\Omega_M + \Omega_\Lambda= 1$ is assumed.
For a different sample of 58 low-redshift and 48 high-redshift SNe~Ia with
luminosity calibrations done using the PRES method
(a generalization of the $\Delta m_{15}$ method),
the results are
$\Omega_M=0.43^{+0.17}_{-0.19}$ and
$\Omega_\Lambda= 1.18^{+0.27}_{-0.28}$ ($1\sigma$ uncertainties)
if no constraints are imposed, or
$\Omega_M = 0.18^{+0.05}_{-0.04}$ if
$\Omega_M + \Omega_\Lambda= 1$ is assumed.

\end{abstract}


\section{Introduction}

\subsection{Distances, Acceleration, and Deceleration}

Studies of the distance versus redshift ($z$) relation of high-redshift 
Type Ia supernovae (SNe~Ia) have produced persuasive evidence that we 
live in a Universe whose expansion rate is currently accelerating.
This remarkable conclusion is the result of developments
in empirical calibration strategies, instrumentation, and software,
which converged to make SNe~Ia very effective tools for estimating 
luminosity distances of very distant galaxies.

The relevant work can be traced back to the early 1990s, when the
light-curve shape versus luminosity relation for SNe~Ia was first
clearly recognized (Phillips 1993).
This breakthrough prompted the building of homogeneous samples of 
SNe~Ia distant enough to be in the Hubble flow.
SNe~Ia were searched for, discovered, carefully observed, and used as
benchmarks for calibrating empirical relationships between light-curve shape
and luminosity (Hamuy et al. 1995, 1996a,b; Riess et al. 1995; 1996, 1998a;
Perlmutter et al. 1997, 1999; Phillips et al. 1999).

In addition to these analytical tools, the measurement of precise luminosity
distances to samples of high-$z$ SNe~Ia required large-format CCDs on 4.0-m
class telescopes, sophisticated software for comparing images taken at 
different epochs, and
efficient human organization to carry out the job.
The first handful of distant SNe~Ia already challenged 
the theoretically preferred cosmological model of the time:
a zero-curvature universe closed by the mass density,
$\Omega_M = 1$ (Garnavich et al. 1998a; Perlmutter \etal\ 1998).

When the distant SN~Ia samples became large enough to reduce the
statistical noise of the combined distance moduli to
a few hundredths of a magnitude, evidence that the Universe is
currently accelerating became clear: at $z \approx 0.5$, the distances
are larger than expected for a universe with any degree of
deceleration (Riess et al. 1998a; Perlmutter et al. 1999).
This result was reached using two different
samples of distant SNe~Ia by two different groups of astronomers and
physicists, the High-$z$ SN Search Team (HZT, led by B. P. Schmidt;
Schmidt et al. 1998), and the Supernova Cosmology Project (SCP, 
led by S. Perlmutter; Perlmutter et al. 1997).
This helped lend credence to the controversial result.
The two groups used the same observables (luminosity
distances and redshifts of high-$z$ SNe~Ia), but
the fact that they had different discovery and analysis tools
provided some resilience against systematic effects.

More recent observations confirm and extend the 1998--1999 findings.
Tonry et al. (2003), Knop et al. (2003), and Barris et al. (2004)
use SN~Ia samples, observed with the {\it Hubble Space 
Telescope (HST)} and ground-based telescopes, to demonstrate 
that the results of Riess et al. (1998a) and Perlmutter et
al. (1999) were not a statistical fluke.
Distances of SNe~Ia at $z \approx 0.5$ are confirmed to be
larger than those predicted for an empty universe.
In addition, Tonry et al. (2003) and Barris et al. (2004)
find that the result is likely due to cosmology, 
not to an unrecognized systematic effect:
at $z \ga 0.7$ the trend appears to be reversed, with
luminosity distances growing with redshift at a rate
{\it slower} than that predicted by an empty universe.
This effect is more convincingly shown by Riess et al. (2001) 
and Riess et al. (2004) with {\it HST} observations of SNe~Ia 
at $z > 1$, strongly suggesting that the Universe was decelerating
at early times.

The widespread interpretation of the combined high-$z$ SN~Ia
observations has been that the Universe
has a ``Cosmological Constant,'' or some other form of ``dark energy''
accelerating its present rate of expansion.
Reviews with more detail on this remarkable paradigm shift are given
by Riess (2000a), Leibundgut (2001), Carroll (2001), Perlmutter
\& Schmidt (2003), and Filippenko (2001, 2004, 2005).

The Hubble diagram of distant SNe~Ia is not the only
experiment suggesting the presence of dark energy and an
accelerating universe.
The best-known method is the combination of observations of the
power spectrum of fluctuations in the cosmic microwave background 
radiation (CMBR) with independent measurements of the mass-density 
parameter from galaxy clusters.
The former favors a Universe with $\Omega_{total} = 1.0 \pm 0.02$
(de Bernardis et al. 2002; Spergel \etal\ 2003), while the latter
indicates that $\Omega_M \approx 0.3$ (Peacock 2001), or perhaps even
lower ($\Omega_M \approx 0.19$, Bahcall et al. 2003).
These results indicate that a significant
contribution from dark energy is necessary to produce the nearly flat
geometry of the Universe, although they do not provide direct
evidence for {\it acceleration}.

Very recently, two other experiments produced results that support
the presence of dark energy.
First, studies of the Integrated Sachs-Wolfe (ISW) effect based on the 
cross-correlation between the Sloan Digital Sky Survey sample of galaxies
and the temperature of the CMBR measured by the Wilkinson Microwave Anisotropy
Probe (Afshordi, Loh, \& Strauss 2004; Boughn \& Crittenden 2004; Fosalba 
et al. 2003; Nolta et al. 2004; Scranton et al. 2005).
The cross-correlation
is achromatic and shows more clearly in the subsamples of galaxies
at larger mean redshifts ($z \approx $0.43, 0.49, and 0.55) for which 
the null hypothesis (i.e., no dark energy) is excluded at $>$90\% confidence.
Although this test provides physical evidence for the existence of dark
energy, it cannot give details on its nature because
the ISW effect is essentially independent of the dark-energy properties.

Second, Allen et al. (2004) use Chandra observations of the largest,
apparently relaxed, galaxy clusters to set joint constraints on the 
dark energy and mass density parameters.
They use the dependence of the derived baryonic mass fraction on the
angular diameter distances to the clusters, plus the {\it assumption} that this
baryonic mass fraction should be constant with redshift for the largest
clusters, to set a test for the reference cosmology:
the baryonic mass fraction will appear to be constant
only if the angular diameter distances are computed with the
correct cosmological model.
They obtain a clear detection of dark energy (greater
than 3$\sigma$), although one may question the assumptions that the
clusters are relaxed and have a constant baryonic mass fraction.

\subsection{The Continuing Test}

The convergence of independent observations to a unified
``concordance'' vision of the Universe --- with low mass
($\Omega_M \approx 0.3$), a large component of dark energy
($\Omega_\Lambda \approx 0.7$, where we use $\Lambda$ 
generically for dark energy), and zero curvature (preserving 
the initial condition favored by inflation) --- has a 
very strong appeal to cosmologists and astrophysicists, and 
the results are now taken as nearly certain.
Nevertheless, the HZT has been committed to continue
testing the result by reducing statistical uncertainties and 
constraining systematic effects with more certainty.

Probing alternative interpretations of the apparently dimmer
than expected SNe~Ia remains important now, but
it was particularly pressing in 1998, when the observational 
campaign reported in this paper was conceived.
Alternatives seriously viewed at the time were as follows.
(a) Some of the high-$z$ SNe had been misclassified
due to low signal-to-noise ratios (S/N) in spectra and light curves;
they were actually subluminous SNe~Ia, or even the fainter SNe~Ib/Ic,
biasing the average derived distances toward larger values.
(b) Some of the high-$z$ SNe~Ia 
had been made dimmer by an anomalous ``grey'' extinction,
caused by dust in the form of carbonated needles instead of 
the typical grains of carbon, iron, and silicates.
(c) The objects
were drawn from a SN~Ia sample statistically different from the nearby
population, making a direct comparison unwarranted.
(d) SNe~Ia
were not standard candles in a cosmological sense, but subject 
to evolutionary effects that made them intrinsically dimmer with 
lookback time.
These issues were pondered by Schmidt at al. (1998), Riess et al. (1998a),
and Perlmutter et al. (1999).
Although there were, at the time, strong reasons to suggest that none of
them implied an effect large enough to nullify the apparent 
acceleration of the Universe,
they were considered important matters of concern that needed to be
experimentally addressed.

Each one of the previous alternative explanations
could be tested through specific observational strategies.
Item (a) required better spectroscopy and photometry of the 
high-$z$ SN candidates, preferably 
over a wavelength range extending to rest-frame infrared. 
For example, one of the clearest differences between normal 
and subluminous SNe~Ia (or SNe~Ib/Ic) is the infrared light 
curve which respectively shows, or does not show, 
a secondary maximum.
Optical spectra having larger S/N, and extending as far to the
red as possible, were also a high priority.
Regarding (b), multicolor photometry extending to infrared
wavelengths was desired.
No ``grey'' dust produces a completely grey extinction 
(Aguirre 1999a,b).
Although the $B-V$ color excess of dust needles is
smaller than observational uncertainties, the color
excess over wider wavelength intervals is larger and may be tested.
Therefore, one of the first priorities after the work of 
Riess et al. (1998a) was to obtain rest-frame $I_c$ photometry of
high-$z$ SNe~Ia, through observations in the 
infrared $J$ band (1.2~$\mu$m).

Item (c) above is difficult to resolve, but probably
has a limited impact because high-$z$ and low-$z$ SNe~Ia are 
not statistically compared as populations but as individuals, 
and each one of them is individually calibrated in luminosity 
using its own light-curve shape.
After correcting for light-curve shape,
the intrinsic dispersion in luminosity of SNe~Ia
is very small (Riess et al. 1995; Hamuy et al. 1996a; Phillips et al. 1999;
Jha 2000; Germany 2001), implying that any residual Malmquist bias
will be minimal (see the discussion in Schmidt et al. 1998).
Contaminants of the sample like SNe~Ic, which have a much broader
distribution in luminosity, might introduce a luminosity bias.
The impact of putative differences in population, both of SNe~Ia 
and contaminants, is reduced simply by increasing the S/N of
the observations, thereby providing better discrimination 
between normal and peculiar SNe and improving the brightness 
calibration of each individual event.
So, a second obvious priority that we set for the campaigns 
following 1998 was to increase the S/N as much as possible, 
if necessary restricting the number of objects to monitor, 
in a ``quality over quantity'' approach.

Finally, evolutionary effects in the light sources, item (d) above,
traditionally pose an insurmountable problem for
any Hubble diagram at high redshift, since detailed knowledge of 
the source evolution with lookback time is required.
In this particular case, however, astronomers have a
simpler solution.
Current estimates of $\Omega_\Lambda$ imply that its dynamic effect 
started to dominate over that of $\Omega_M$ a few billion years ago, 
with the redshift of the transition being $z_T \approx 0.5$ (Riess et 
al. 2004).
At redshift larger than $z_T$, luminosity distances in a low mass,
decelerating (at that time) universe should grow more {\it slowly} 
with increasing redshift than in an empty universe.
By observing SNe beyond $z_T \approx 0.5$, the issue of 
evolutionary (and other systematic)
effects can be experimentally tested.

The observational campaigns of the HZT from 1999 onwards were
designed with these concerns and strategies in mind.
The first HZT campaign of the year 1999, the one reported
here, was designed to address items (a), (b), and (c) above.
Its goal was to obtain superior-quality light curves and spectra 
of a few SNe at a redshift of $\sim$0.5 to provide a point in the
Hubble diagram with high statistical
significance at this critical redshift.

The SN search was programmed at Cerro Tololo Inter-American 
Observatory (CTIO), and spectroscopic confirmation was done at
the Keck-1 10~m telescope.
Follow-up observations were conducted at several ground-based 
telescopes and with the {\it Hubble Space Telescope (HST)}.
Telescopes that delivered images useful for the SN light curves 
presented here were the Keck-1 and Keck-2 telescopes (Keck Observatory), 
VLT-1 (ESO-Paranal Observatory), the 3.5~m telescope at Apache 
Point Observatory, the 3.6~m telescope at ESO-La Silla Observatory, 
the 2.2~m telescope of the University of Hawaii (Mauna Kea 
Observatory), the 4.0~m and 1.5~m telescopes at CTIO, and {\it HST}.

As a side project of this campaign, we observed SN~1999Q in the
$J$ band.
We transformed the observed $J$ data into rest-frame $I_c$, 
studying the infrared properties of this SN and searching for 
the color signature of relatively ``grey'' reddening by dust.
These results were published by Riess et al. (2000b).
Briefly, we found that SN~1999Q was a normal SN~Ia, and there
was no evidence for non-standard extinction.

In \S~\ref{se:observations} we describe our search campaign, 
the follow-up spectroscopic and photometric observations, and 
the calibration of the local sequences of standard stars.
We present and describe the observed-frame SN photometry in
\S~\ref{s:SN_phot}.
Section \ref{se:SN_rest} describes the transformation of the 
observed-frame photometry into the rest frame of each SN, 
together with the methods of light-curve fitting, K-correction, 
and luminosity calibration.
We analyze and discuss the results
in \S~\ref{se:discussion}, presenting updated
high-$z$ Hubble diagrams and using them to fit cosmological 
parameters.

\section{Observations} \label{se:observations}

\subsection{Search, Discovery, and Confirmation} \label{ss:search}

Three nights with the 4.0~m Blanco telescope at CTIO had been 
allocated for this search: 1999 January 13, 14, and 17 (UT dates
are used throughout this paper).
The telescope was equipped with the ``Big Throughput Camera''
(BTC; Wittman et al. 1998), a prime focus
mosaic imager with four SITe $2048 \times 2048$ pixel CCDs,
a field of view of 0.24 square degrees, 
and a pixel scale of 0.43\arcsec\ pix$^{-1}$.

No time for first-epoch observations could be scheduled during 
the previous lunation.
Instead, the image comparison to detect transient sources was 
done with templates obtained during the campaigns of 1998 
(Suntzeff et al. 2006, in preparation), approximately a year 
earlier.
This made the search vulnerable to objects such as quasars, which 
vary on a time scale longer than SNe~Ia near maximum brightness.
The search was also prone to the discovery of SNe many days after 
maximum, when their utility as distance estimators is very limited.
Our strategy to refine the sample before investing
spectroscopic time was to search for SN candidates as rapidly as 
possible, repeat images of the candidate fields during the nights 
between discovery and spectroscopy, and apply photometric cuts 
based on preliminary light-curve evolution and colors of the 
transient sources.
Additional criteria included the presence of a suspected 
host galaxy, and the projected distance of the transient source from the 
galaxy nucleus.

The pipeline used for image reduction and analysis is described
by Schmidt et al. (1998).
Briefly, it starts at the mountaintop, where the BTC images are 
trimmed, bias-corrected, flat-fielded, and defringed (if necessary) 
with standard IRAF\footnote{The Image Reduction and 
Analysis Facility is developed and maintained by NOAO, under 
contract with the National Science Foundation.} tasks.
Images are then transferred to a network of workstations at the 
CTIO headquarters in La Serena, where they are aligned, matched 
in point-spread function (PSF) shape and
intensity with the template frames, and subtracted.
The residual images are searched using a PSF detection algorithm,
and a list of candidate transient sources is generated for human 
experts to inspect by eye.

The search and photometric confirmation of candidates at CTIO was 
very successful, providing 37 transient sources deemed sufficiently
good for subsequent spectroscopy.
Of these, 14 candidates were given first priority and
another 14, second.
Priorities were assigned according to apparent brightness, change 
of brightness between different epochs (when available), color, 
presence of a putative host galaxy, and position of the candidate 
with respect to the galaxy nucleus.
The 9 remaining events were considered true transient 
stellar-like sources.
However, they had a combination of brightness, color, and absence 
of a host galaxy that made them unlikely to be SNe useful for our 
science goals, and were thus dropped from further consideration.

\subsection{Keck-1 Spectroscopy, SN Classification, and 
Selection} \label{ss:spectroscopy}

Spectroscopy of the selected SN candidates was done with the Low 
Resolution Imaging Spectrometer (LRIS; Oke et al. 1995) at the
Cassegrain focus of the Keck-1 telescope.
Five nights (1999 January 17--21) had been allocated to our program.
Skies were clear most of the time, with only three hours of the 
last night lost to fog.
The atmospheric seeing was mediocre, however, with a median value 
close to $2''$ and typical variations between $1.2''$ and $2.5''$.
Spectra were obtained in long-slit mode, with a slit width of 
$1.0''$ or $1.5''$ depending on the seeing.
The spectrograph was rotated to include in the same slit the 
SN candidate and a nearby star.

Reductions of the two-dimensional (2D) spectra were done following 
the typical procedures (e.g., Matheson et al. 2005) and using
IRAF tasks.
The bias level was removed using the overscan regions, and the
2D spectra were flat-fielded with dome flats.
The location of the nearby star on the detector as a function of
position along the dispersion was used to define a trace, which 
was then shifted and centered on the SN candidate to improve the
quality of the extraction. 
Care was taken to model the light of the background sky, including 
the host galaxy when present, to minimize contamination of the SN 
candidate spectrum.
Since the typical angular size of the surface brightness variations 
in these fairly high-redshift galaxies is similar to both the
slit width and the seeing, it is impossible to 
completely remove the background and some contamination remains.

Despite the mediocre seeing, the Keck-1 observers managed to
take spectra of the 28 candidates of first and second priority.
One half of them were recognized as SNe: nine were classified
as SNe~Ia, four as SNe~II, and one as a SN~Ic.
Five additional candidates were quasars.
Nine of the spectra could not be assigned to any kind of known 
object.
In three cases, this happened even though the spectra had
decent S/N.

\begin{figure}[t]
\plotone{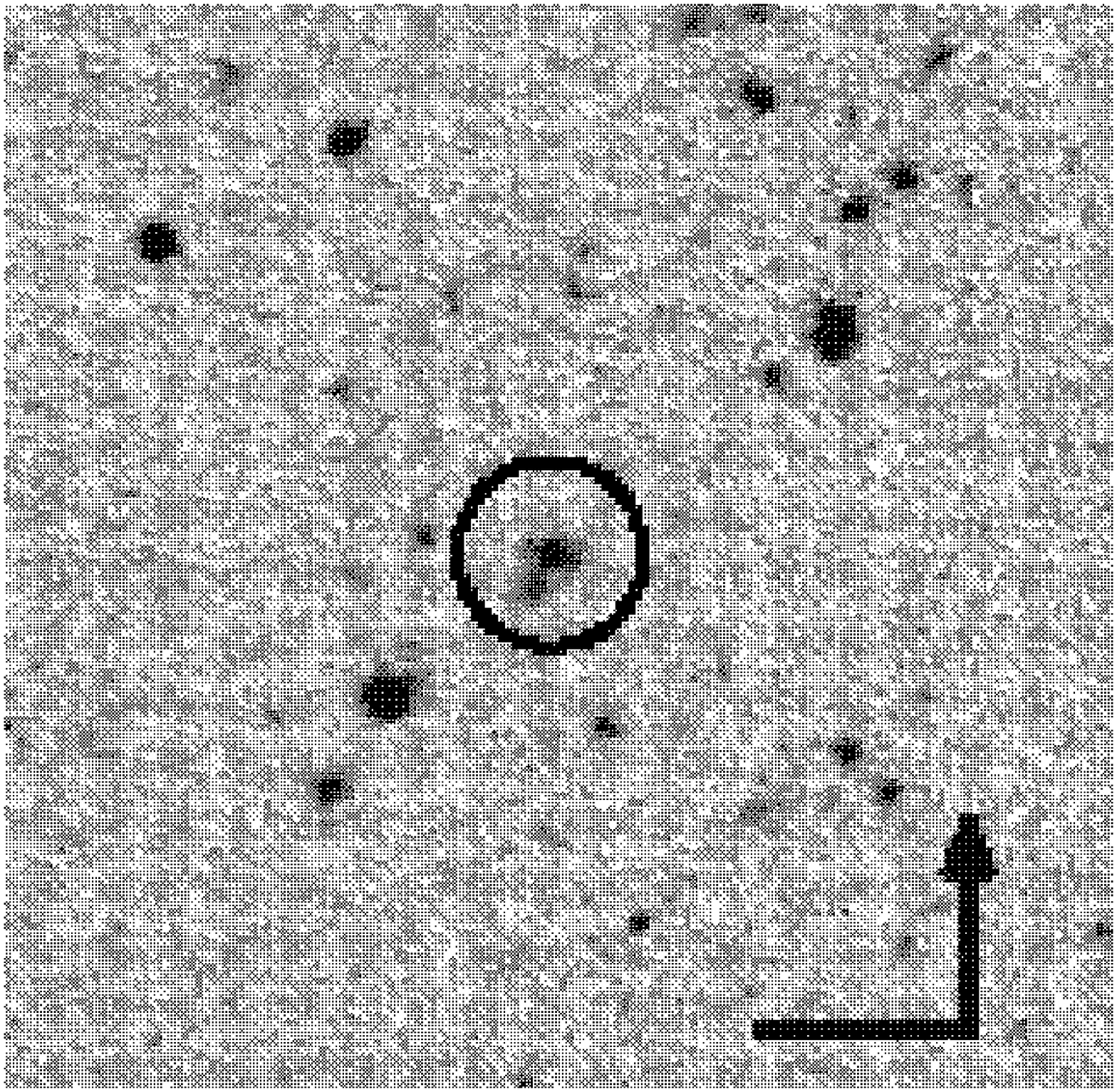}
\caption{Finder chart for SN 1999M. 
This is the discovery
image taken with the CTIO 4.0~m telescope on 1999 Jan. 13,
through an $R_c$ filter.
SN 1999M is centered in the circle.
The width of the image shown is $\sim$66$''$.
North (up) and east (left) are indicated by the arrow 
and line, respectively.
\label{fi:SN99M_finderchart}}
\end{figure}

\begin{figure}[t]
\plotone{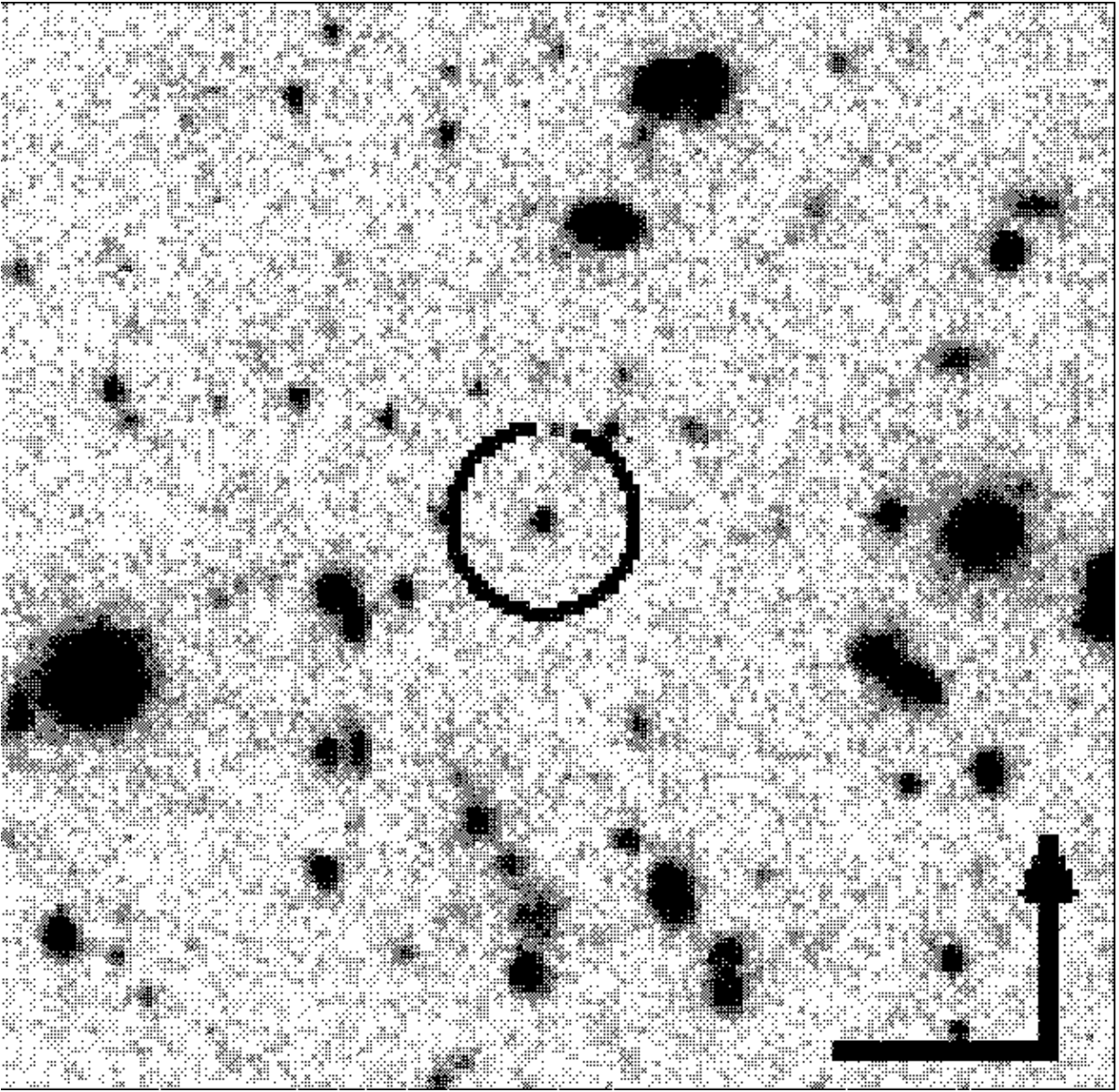}
\caption{Finder chart for SN 1999N, obtained on 1999 Jan. 13.
See Figure 1 for other details.
\label{fi:SN99N_finderchart}}
\end{figure}

\begin{figure}[t]
\plotone{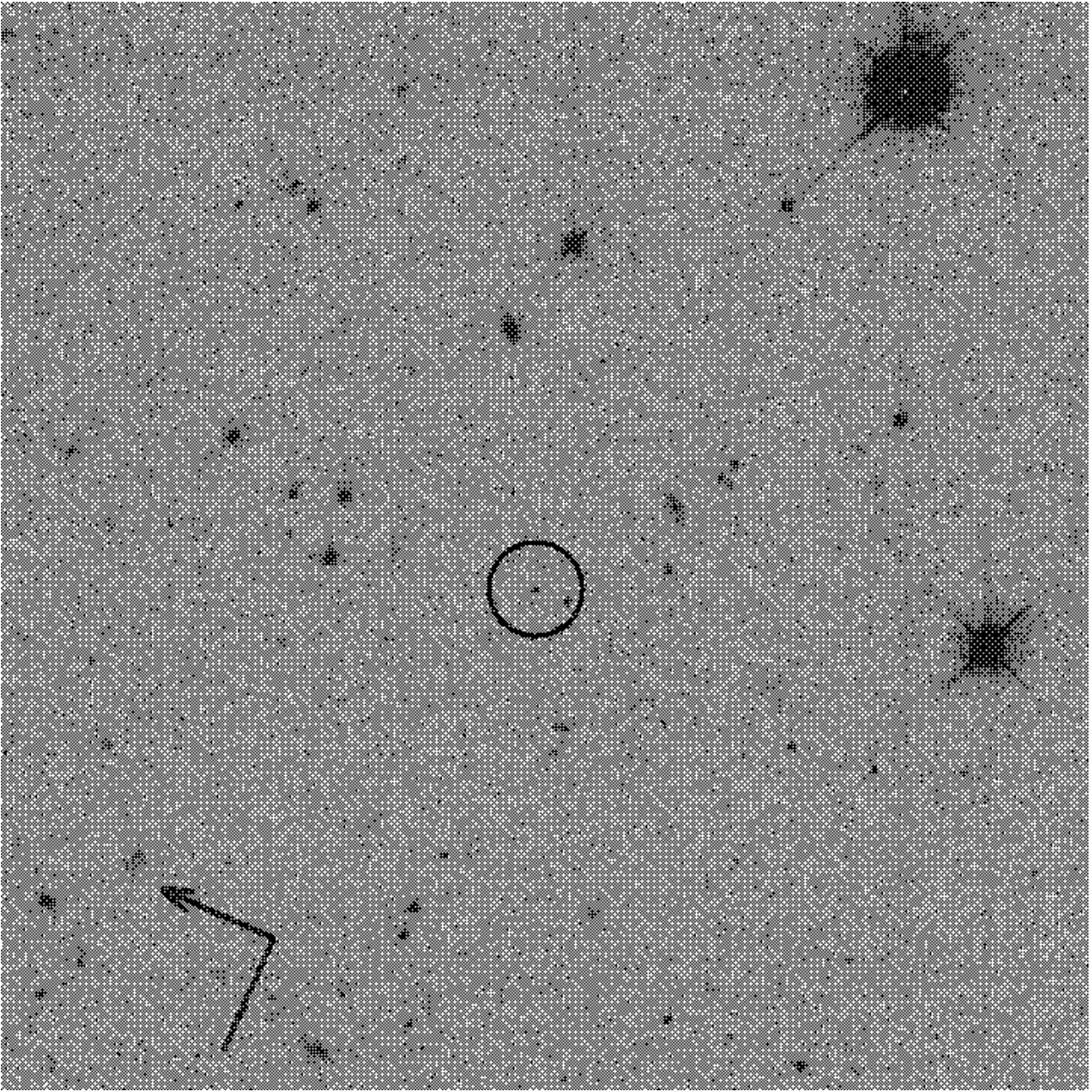}
\caption{Finder chart for SN 1999Q.
The image is a combination of 6 {\it HST} WFPC2 images taken through filter F675W,
between 1999 Feb. 1 and 1999 Mar. 7.
SN 1999Q is centered in the circle.
The width of the image shown is $\sim$66$''$.
North and east are indicated by the arrow 
and line, respectively, near the bottom-left corner of the image.
\label{fi:SN99Q_finderchart}}
\end{figure}

\begin{figure}[t]
\plotone{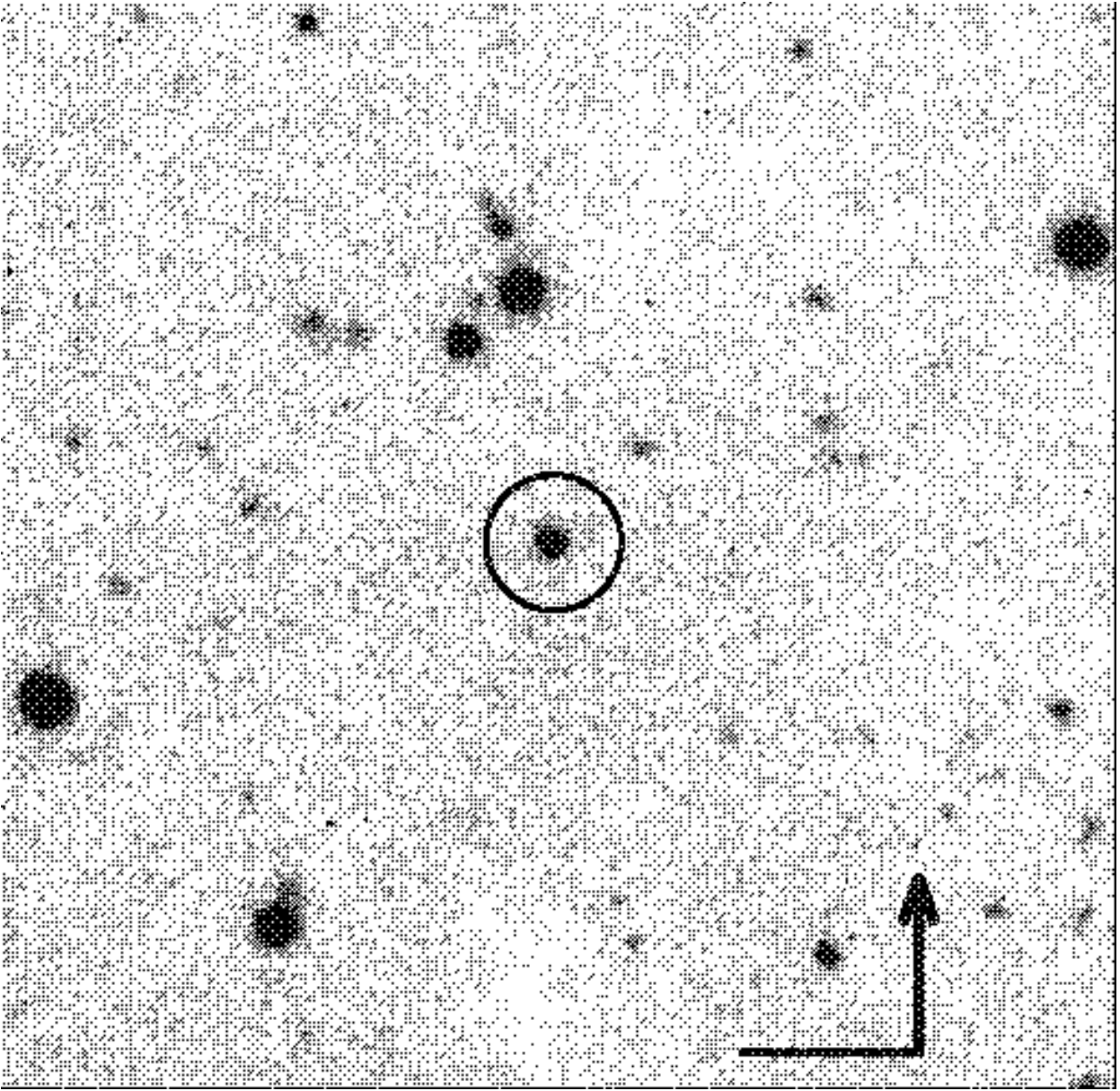}
\caption{Finder chart for SN 1999S.
This $R$-band image was obtained on 1999 Feb. 9 with LRIS on the
Keck-2 telescope.
SN 1999S is centered in the circle.
The width of the image shown is $\sim$66$''$.
North (up) and east (left) are indicated by the arrow 
and line, respectively.
\label{fi:SN99S_finderchart}}
\end{figure}

\begin{figure}[t]
\plotone{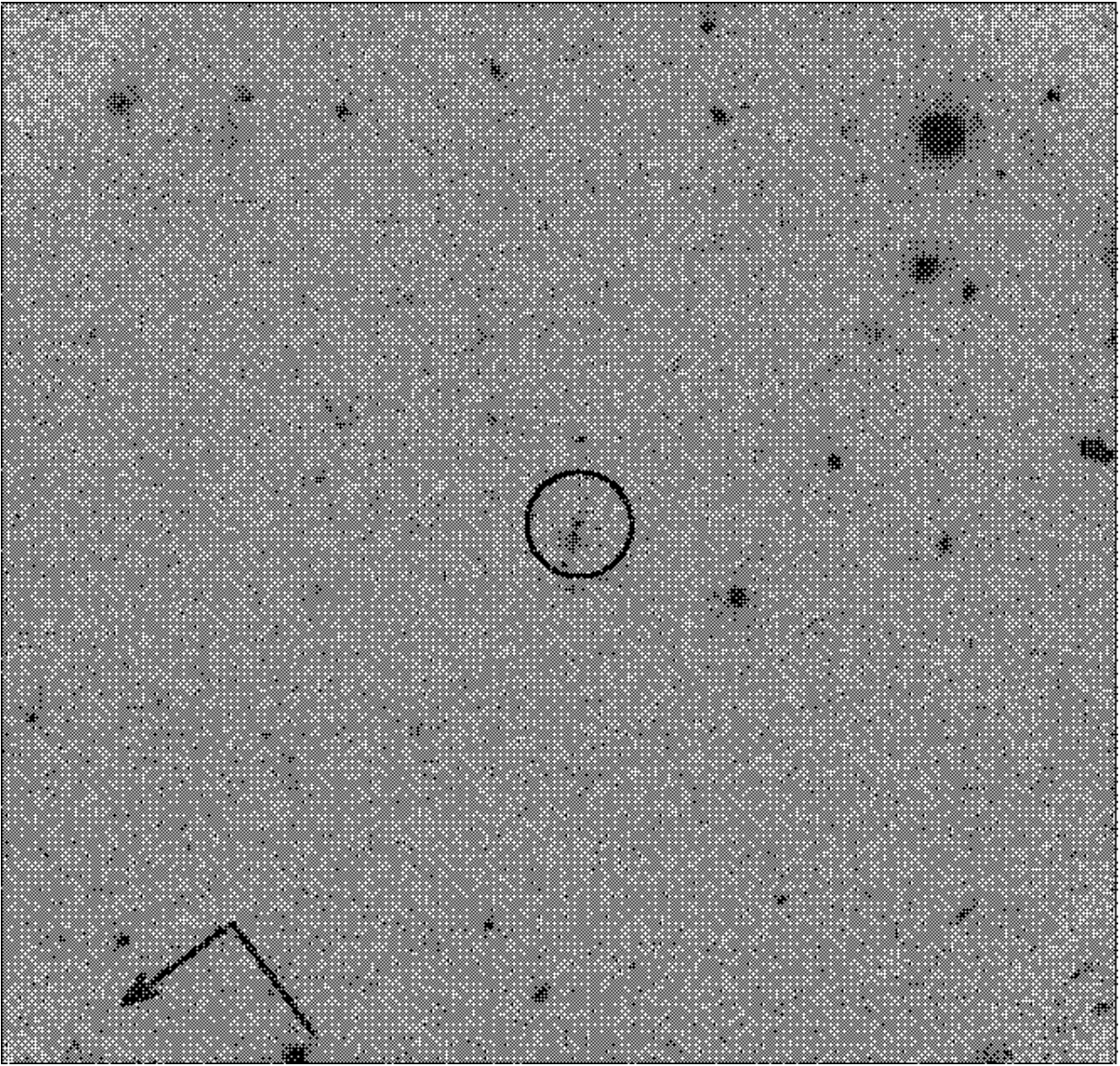}
\caption{Finding chart for SN 1999U.
The picture shows a combination of 6 {\it HST} WFPC2 images taken through filter F675W,
between 1999 Feb. 1 and 1999 Mar. 7.
See Figure 3 for other details.
\label{fi:SN99U_finderchart}}
\end{figure}

Five of the nine SNe~Ia were selected for follow-up observations,
including {\it HST} imaging for two of them.
Selection of the SNe was based on a combination of their redshift,
their initially assigned priority (which essentially grades their 
quality with photometric criteria), and their spectroscopic age.
SNe~Ia at $z \approx 0.5$, with normal colors, and not very close 
to the nuclei of bright host galaxies were given the highest priority 
for follow-up observations.
Brief information on all the spectroscopically confirmed candidates
was provided by Garnavich et al. (1999).
Figures \ref{fi:SN99M_finderchart} to \ref{fi:SN99U_finderchart} give
finder charts of the five SNe~Ia analyzed in this paper.
Note that the pixelization of the ground-based and {\it HST} images 
differs, for approximately the same field of view.

Table \ref{ta:candidates} summarizes additional information on 
the chosen SNe~Ia.
We give their position in the sky, the foreground Galactic 
extinction from Schlegel et al. (1998), and the total number of
images accumulated from various telescopes, including the 
very late-time images obtained when the SNe had faded away.
The table includes all the images that were collected in this 
study's database, even though some of them were subsequently
combined into one, and some others were discarded due to low 
S/N or very poor seeing.
The photometric follow-up campaign resulted in more than 300 
images taken with nine different telescopes.
The total number of images for each SN is a good
indication of its assigned priority.

\begin{figure}[t]
\includegraphics[scale=0.65, angle=270.0 ]{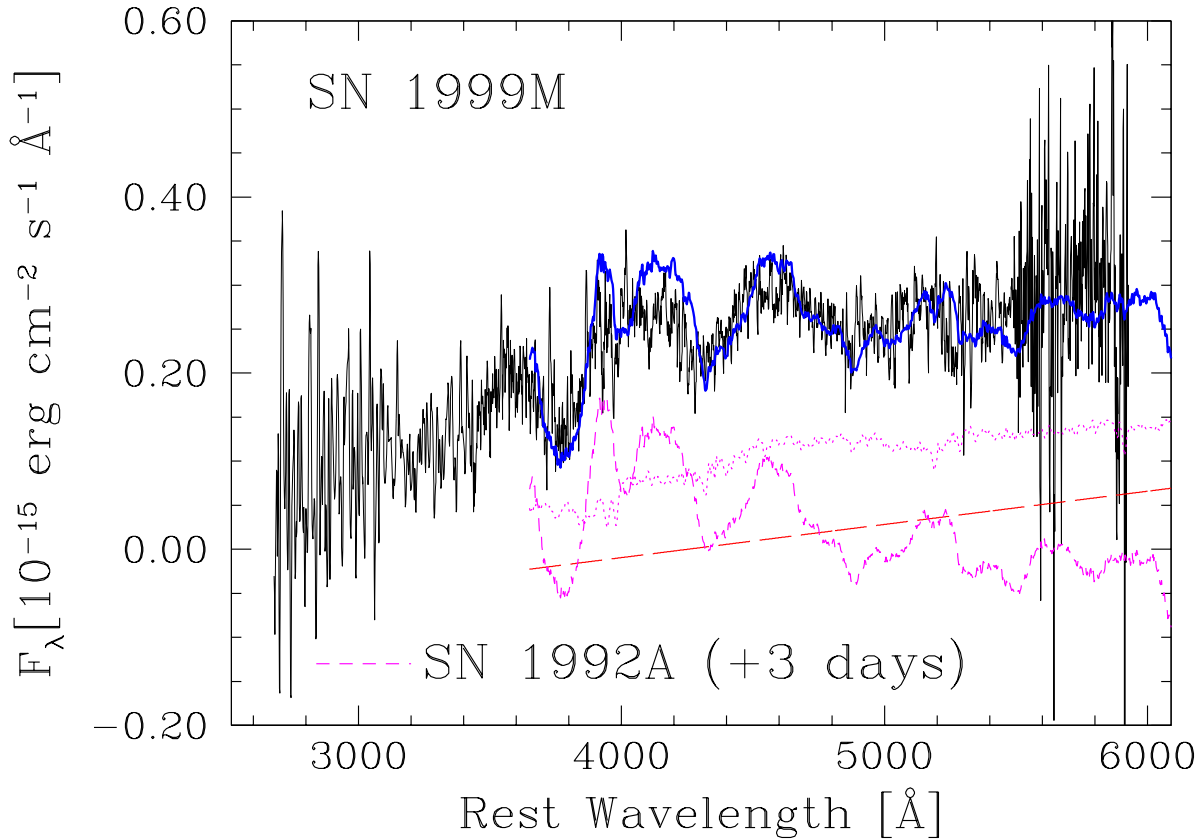}
\caption{Keck-1 spectrum of SN 1999M (thin solid line).
The long dashed line shows the unknown slope (see text).
The short dashed line give the nearby SN spectrum taken as a model.
The dotted line is the spectrum of the contaminating galaxy.
The SN model spectrum is SN~1992A, 3 days after
$B$ maximum (Kirshner et al. 1993).
The thick solid line is the combination of
the nearby SN, galaxy, and slope.
All spectra have the units indicated along the
ordinate, with the exception of the isolated
spectrum of SN 1992A, which has been shifted
by an additive constant for clarity.
\label{fi:spec_sn99M}}
\end{figure}

\begin{figure}[t]
\includegraphics[scale=0.65, angle=270.0 ]{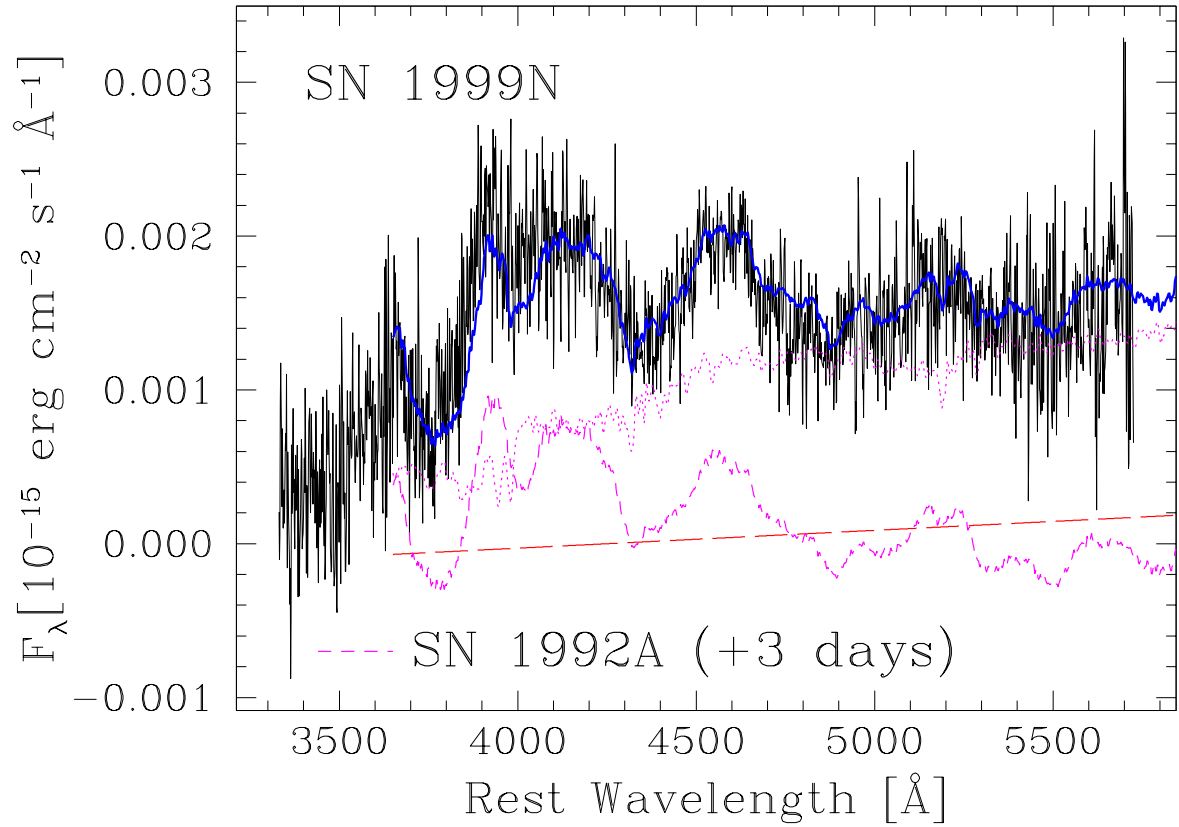}
\caption{Keck-1 spectrum of SN 1999N.
Elements are the same as in
Figure \protect{\ref{fi:spec_sn99M}}, although
the flux scale is different.
\label{fi:spec_sn99N}}
\end{figure}

\begin{figure}[t]
\includegraphics[scale=0.65, angle=270.0 ]{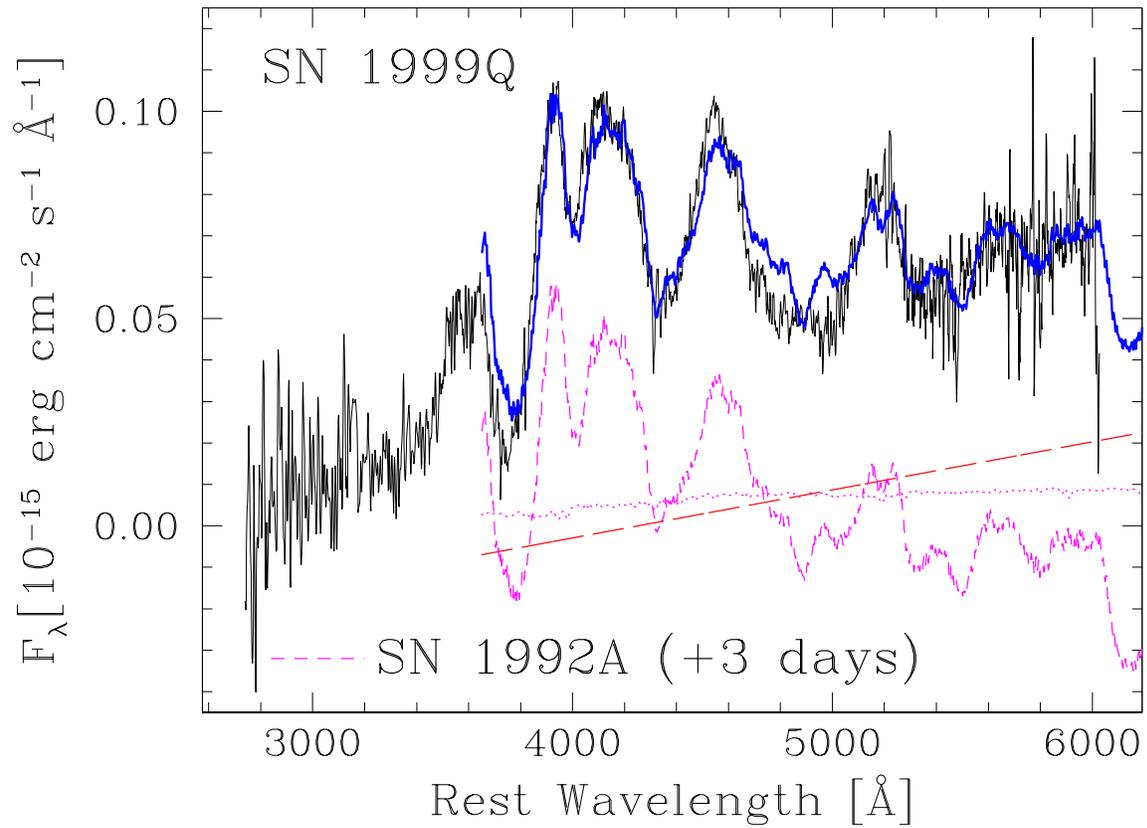}
\caption{Keck-1 spectrum of SN 1999Q.
Elements are the same as in
Figure \protect{\ref{fi:spec_sn99M}}, although
the flux scale is different. 
Of the SNe in our sample,
this one is the least
contaminated by the host galaxy.
\label{fi:spec_sn99Q}}
\end{figure}

\begin{figure}[t]
\includegraphics[scale=0.65, angle=270.0 ]{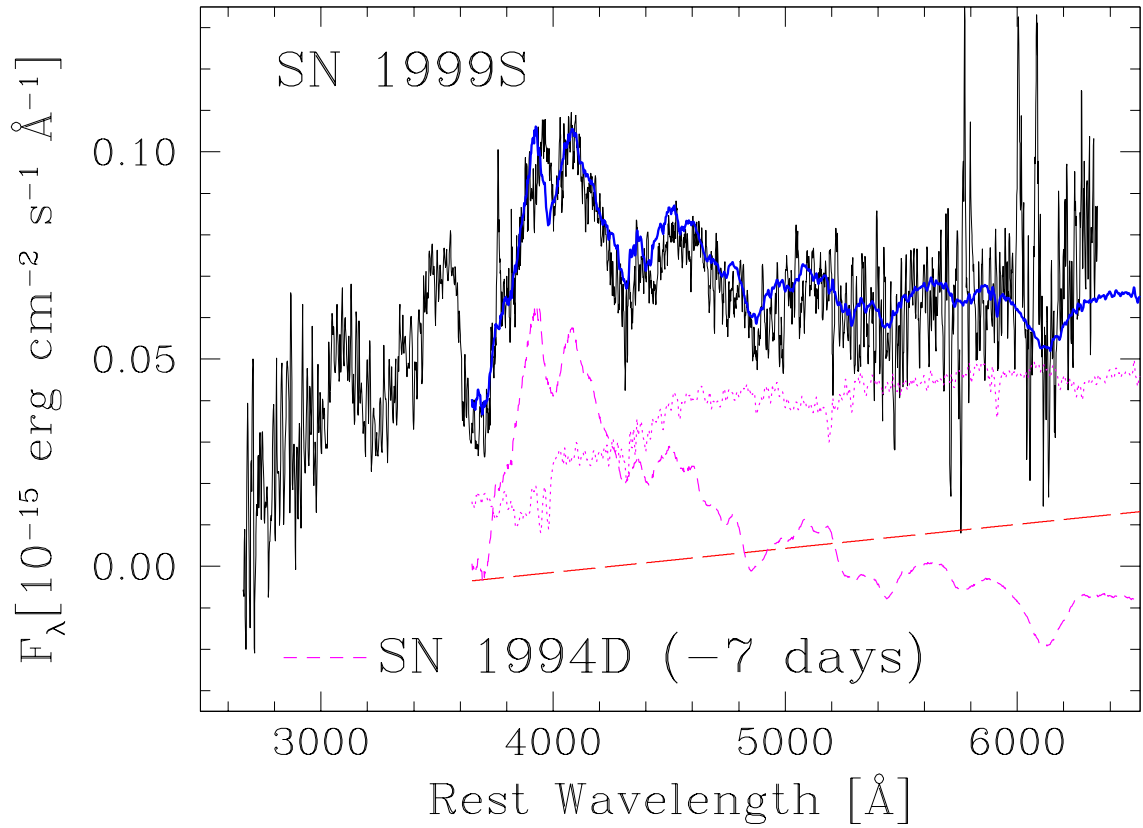}
\caption{Keck-1 spectrum of SN 1999S.
Elements are the same as in
Figure \protect{\ref{fi:spec_sn99M}}, although
the flux scale is different. 
The SN model spectrum
in this case is SN~1994D, 7 days before $B$ maximum
(Richmond et al. 1995).
\label{fi:spec_sn99S}}
\end{figure}

\begin{figure}[t]
\includegraphics[scale=0.65, angle=270.0 ]{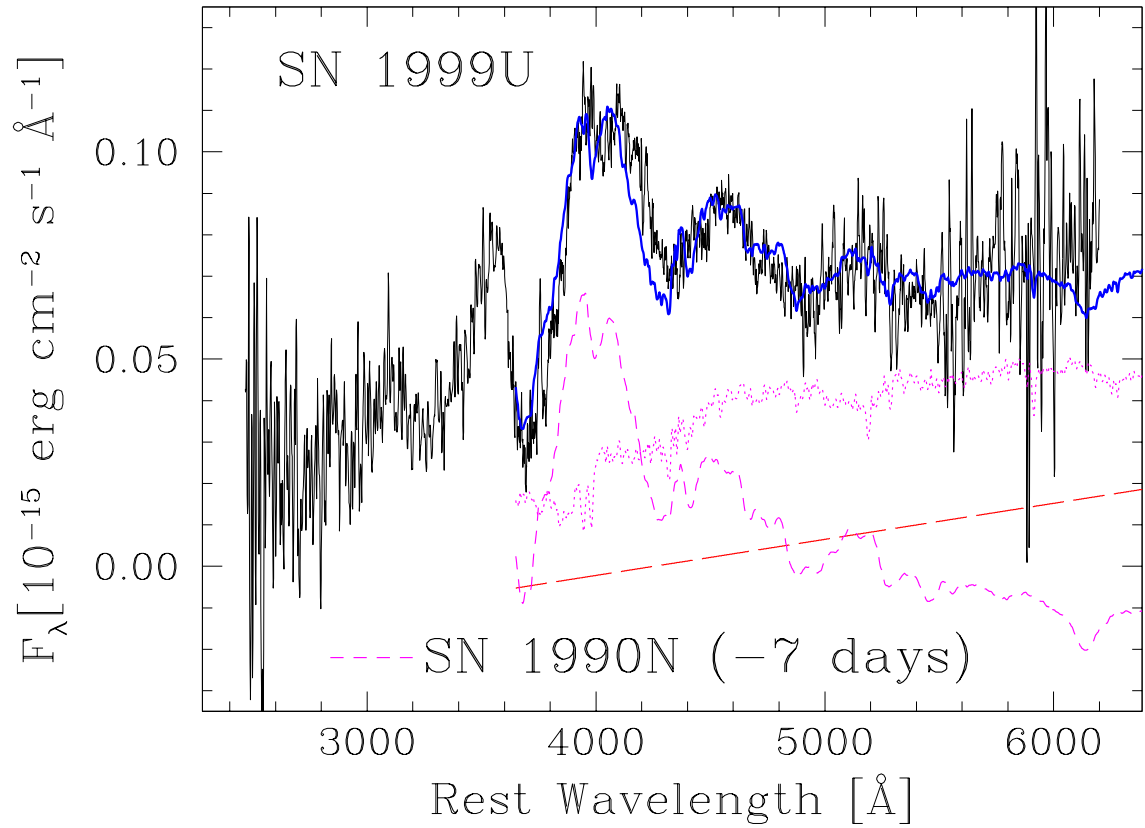}
\caption{Keck-1 spectrum of SN 1999U.
Elements are the same as in
figure \protect{\ref{fi:spec_sn99M}}, although
the flux scale is different. 
The SN model spectrum
in this case is SN~1990N, 7 days before $B$ maximum
(Leibundgut et al. 1991).
\label{fi:spec_sn99U}}
\end{figure}

The concentration of spectroscopic time on just a few 
candidates resulted in high S/N. 
It ranges from $\sim$15 for
SN~1999Q, our highest-priority object, down to $\sim$6 for
SN~1999M and SN~1999N.
The latter two had been given low priority because of the 
bright background of their host galaxies.
Figures \ref{fi:spec_sn99M} to \ref{fi:spec_sn99U} show the 
final Keck-1 spectra.
For comparison, we also include spectra of nearby SNe~Ia at
early phases combined with the spectrum of a galaxy from
the catalog of Kennicutt (1992).

Table \ref{ta:spectra_data} gives additional details of the 
spectroscopy: the method used to obtain the redshift of each SN,
the approximate phase of the spectra obtained from the 
spectral-feature age method (see Riess et al. 1998b), and the 
phase implied by two different light-curve fits to be described 
in \S~\ref{se:SN_rest}.
The redshifts given in Table~\ref{ta:spectra_data} are slightly 
different from those initially reported after a quick analysis 
during the search (Garnavich et al. 1999), and supersede them.

It is clear from the figures that all the objects selected
are SNe~Ia without obvious spectroscopic peculiarities, and 
that they were detected at early phases.
Near maximum light, typical SNe~Ia show a strong \ion{Si}{2} 
absorption line at rest-frame wavelength $\sim$615~nm (the 
strongest \ion{Si}{2} line in the optical/near-IR range).
Although the redshift of the objects places this line out of
the CCD sensitivity window, all spectra display the second-strongest
silicon line in this range, \ion{Si}{2} $\lambda 4130$, which
appears close to rest-frame 400~nm.
This confirms that our SNe are not Type~Ib/Ic events, for which 
this secondary \ion{Si}{2} line is essentially absent 
(Clocchiatti et al. 2000; Coil et al. 2000).

Figures \ref{fi:spec_sn99M} to \ref{fi:spec_sn99U}, in addition,
detail the basic elements used to match the observed spectra.
These elements are the spectrum of a nearby SN~Ia at early phases and 
the spectrum of a galaxy from the catalog of Kennicutt (1992), but we 
found it necessary to add an additional slope to make its continuum 
redder.
All of the observed spectra appear too red to match the model of a SN
contaminated by a galaxy.
It is possible to make the combined SN-plus-galaxy spectra redder by
resorting to galaxies with extremely red spectra, or by
increasing the fraction of light from the galaxy in the combination.
There is, however, a limit to the amount of light from a galaxy that
the combination can tolerate since the contrast between maxima and
minima of the SN features is changed by galaxy contamination.
Also, there are cases were the continuum of a red galaxy improves
the contrast between the red and blue spectral regions (i.e. the color
of the model), but does not improve the overall fit.
A straight line, as shown in the figures, simply does a better job.

The case of SN 1999Q is especially illustrative.
All the information obtained on this event, including the very deep image
obtained by combining the {\it HST} exposures, indicates that its host
galaxy has very low luminosity.
The spectrum in Figure \ref{fi:spec_sn99Q} confirms this.
The contribution from a galaxy spectrum is very small,
at most around 20\%, while the need for a smooth trend to make the
spectrum redder than that of SN 1992A is apparent.

There is a somewhat independent way of testing this conclusion.
For two of the objects (SN 1999S and 1999U) it is possible to
compute a synthetic color from the spectrum and compare it with 
the observed photometry.
For both SNe the synthetic $R_c - I_c$ result is $\sim 0.5$ mag,
about half a magnitude redder than the
observed $R_c - I_c$ color close to maximum light,
the time at which the spectra were taken
(cf. Tables \ref{ta:data99S} and \ref{ta:data99U}).
The spectra used to compute the synthetic colors, however, include the unknown
contribution from the host galaxy, which is likely to be redder than
a SN~Ia near maximum light.
Red galaxies, like the one used to model the observed spectra in figures
\ref{fi:spec_sn99M} to \ref{fi:spec_sn99U}, at $z \approx 0.5$ will
have observed $R_c - I_c \approx 1$ mag.
It is possible to combine light from a red galaxy and a blue
SN to modify the observed SN color ($R_c - I_c \approx 0.0$ mag) and make it
similar to the synthetic color ($R_c - I_c \approx 0.5$ mag), but unrealistic
parameters are required to do so.
One would need, for example, more than 50\% of the light in the slit
to be from the galaxy, {\em and} that the spectrum of the galaxy be 50\%
brighter than that of the SN in the observed $I_c$ band (rest-frame $V$).
A bluer galaxy will accomplish the same thing with more light contamination, or
an even redder galaxy will suffice with a smaller contamination.
It is unrealistic to think that these kinds of extreme combinations are
needed to fit both cases, especially since both spectra exhibit
[O~II] $\lambda3727$ emission (a line not associated with
particularly red galaxies).

We have convinced ourselves that the spurious slopes are not intrinsic to
the spectra, but unfortunately we cannot offer a reasonable explanation.
Since it appears to be independent of the amount of galaxy background
and color, which differ among the SN spectra,
it is probably an instrumental, or reduction, effect.
Its existence introduces a free parameter in the spectroscopic
match, and our attempt to keep this slope to a minimum
biases the spectra of the background galaxy to the red:
spectra of red galaxies match our observations better
than spectra of bluer galaxies.
All the spectra shown were fitted using as a background model
the spectrum of NGC~3379, one of the reddest galaxies in the
sample of Kennicutt (1992).

Phases (i.e., SN ages) obtained from background-contaminated 
SN~Ia spectra are more uncertain than those obtained from the
spectra of well-isolated SNe~Ia (Riess et al. 1997).
When the background-to-SN light ratio increases,
the contrast between maxima and minima of SN features
diminishes, mimicking a younger SN.
Furthermore, some of the observed spectra shown here
are the combination of exposures taken over the
span of up to five nights ($\sim 2.7$ rest-frame days).
At early phases, the spectrum of a SN~Ia can vary 
significantly on timescales as short as 1--2 days.
Hence, some of our spectra should not be considered as
being representative of a single phase, but rather as 
a time average, and time averaging during the photospheric 
phase changes the contrast of SN spectral features.
Both the background contamination and the time averaging
increase the uncertainty of the spectroscopic phase estimate.
The spectrum comparisons shown in figures
\ref{fi:spec_sn99M} to \ref{fi:spec_sn99U} are thus 
somewhat qualitative.
They are approximately correct, but given the large number of 
essentially free and degenerate parameters involved in the 
matching problem, we do not claim that they necessarily 
correspond to the best fit of an objective minimization scheme.

\subsection{Calibration of Photometry} \label{ss:calibration}

\subsubsection{Ground-Based Images}

We measured the brightnesses of the SNe in ground-based
images by computing relative magnitudes with respect to a 
local standard sequence.
The calibration of these sequences
was done entirely with the CTIO 1.5~m telescope.
Seven nights were granted for this in mid-February 1999.
The telescope was equipped with the Cassegrain focus CCD camera.
Two different focal ratios were used, $f$/18 and $f$/13,
giving a scale of $0.39''$ pixel$^{-1}$ and $0.28''$ pixel$^{-1}$, 
respectively.

During the calibration nights, repeated images of the SN fields together 
with images of Landolt (1992) standard-star fields were obtained.
The 1.5~m telescope was equipped with the typical passbands used by the
HZT. These included the $R_c$ and $I_c$ bands of the
Kron/Cousins system (Kron \& Smith 1951; Cousins 1976), and the
custom bands $B45$ and $V45$ --- the special filters designed by the
HZT that correspond to the Johnson (1955) $B$ and $V$
bands (respectively) at $z = 0.45$ (Schmidt et al. 1998).
During four of the seven nights, images in both filter sets were taken.
Each photometric system was treated as a separate set and their images
considered independently.
The calibration campaign resulted in $\sim$640 additional images to analyze.

All the images from the calibration nights were reduced in the standard
way within the IRAF environment.
The electronic pedestal was subtracted using the overscan area of the
CCD, and the bias images were checked for a 2-dimensional residual 
pattern that the overscan correction could not remove.
Although this residual was consistent with no signal, it was in each
case subtracted anyway because its noise level was also minimal.
Dome flats were used to normalize the pixel-to-pixel sensitivity
variations.
Image airmasses were corrected to their effective value using the
task SETAIRMAS of the IRAF package DAOPHOT.

For the images containing the standard-star fields, we
recognized the fields by eye, checked the profiles of
the standard stars for signs of cosmic rays, CCD defects,
or saturation, and did aperture photometry of those that 
did not show problems.
The adopted apertures ($7''$--$12''$) varied with the typical 
seeing of the images of each set.
Photometry was done using DAOPHOT tasks in IRAF.

Once the photometry of the standard stars was done, we fitted the
parameters for the filter set using equations of the form
\begin{equation}\label{eq:calibration}
m_R = R + Z_R  + \kappa_R X_R + C^1_{RI} (R-I) + C^2_{RI} X_R (R-I),
\end{equation}
where $m_R$ is the observed magnitude in the $R$ band, $R$ is the
magnitude of the star in the standard system, $Z_R$ is the $R$-band 
zero point, $\kappa_R$ is the extinction coefficient in the $R$ band,
$X_R$ is the airmass for the $R$-band observation, $C^1_{RI}$ is the
color term in $R-I$, and $C^2_{RI}$ is the cross term in $R-I$.
This was done for each of the nights, and for each of the 
photometric systems observed.
Ultimately, the seven calibration nights gave rise to
eleven different photometric calibrations.

Five of the seven nights provided results that were consistent
within the uncertainties, as well as with the typical values at
the CTIO 1.5~m telescope, and were considered photometric.
For the other two nights, either the extinction coefficients or 
the color terms (or both) were significantly different, 
prompting us to eliminate them from further consideration.
For all the photometric nights, and the two photometric systems, the
cross terms ($C^2_{RI}$ fitted in equation \ref{eq:calibration}) were
consistent with zero and were set to zero for the definitive fit.

The images of the SN fields were searched for stellar objects using
SExtractor (Bertin 1996), and the list of stars provided 
by the program was refined by eye.
We selected fairly extensive sets of local standards, typically 
between $\sim$20 and $\sim$40, because of the need to calibrate 
images taken with very different parameters (e.g., CCD fields
of view) at different telescopes.
The brightnesses of local standards were measured with
PSF-fitting photometry using the IRAF/DAOPHOT
tasks and procedures.
The PSF photometry of the local standards was usually calibrated 
with small apertures, because some of these standards were rather
faint or had relatively close apparent companions.
The transformation of this photometry to the larger apertures 
of the photometric system standards was done with aperture 
corrections defined by bright, well-isolated local standards.
With the instrumental magnitudes in hand, we inverted the
systems of equations like Eq. (\ref{eq:calibration}), and obtained
calibrated magnitudes for the sequences of local standards.

The extensive sets of local standards proved to be very useful.
By calibrating local standards with a wide color range,
it is possible to check whether color terms are important for
the different combinations of telescopes and instruments used.
Moreover, if necessary, the color terms can be estimated from the
SN frames, without a set of all-sky photometric standards observed
with the same combination of telescopes and instruments.
Finally, by having local standards with a wide range of brightnesses, 
it is possible to calibrate images with very different depths, like those
obtained in this campaign where 1.5~m to 10~m telescopes were used.

\subsubsection{HST Images}

We calibrated the SN brightnesses in the {\it HST} images independent
of any local sequence of standard stars.
The {\it HST} images were reduced using the ``On-The-Fly Reprocessing''
image reduction pipeline, provided by the Space Telescope Scientific
Institute (STScI).
Images had been taken in pairs to eliminate cosmic rays with the
task CRREJ of the IRAF package STSDAS,
which uses the two observed images combined with
knowledge of the noise characteristics of the {\it HST} WFPC2 to 
remove cosmic rays.
PSF-fitting photometry of the stars in the {\it HST} frames
was done using HSTPhot (Dolphin 2000),
which is calibrated with the zero points of Holtzman et al. (1995), by normalizing
the PSF magnitude to aperture photometry with $0.5''$ radius.

\begin{figure}[t]
\plotone{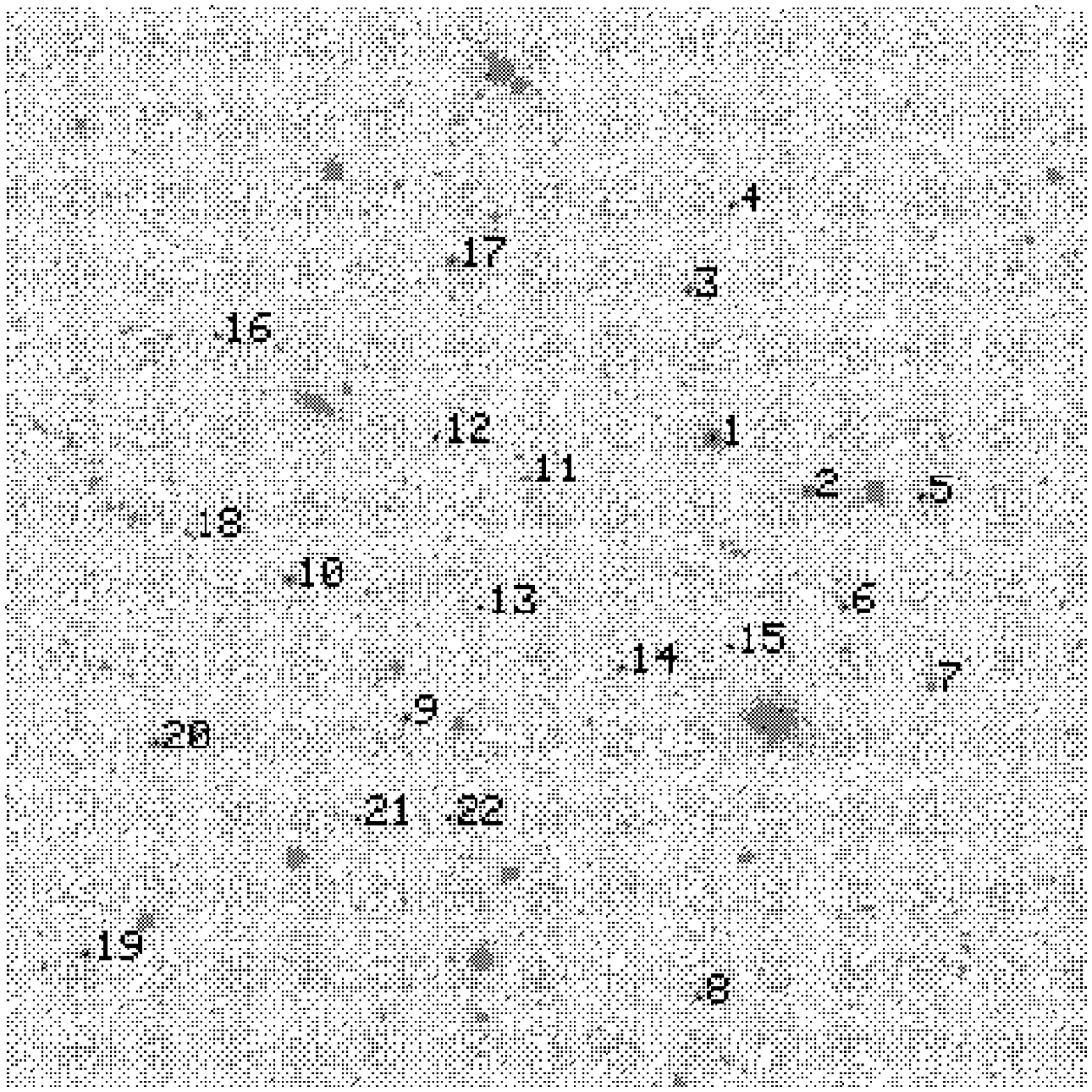}
\caption{Sequence of local standard stars in the field of SN~1999M.
North is up, east is to the left.
The field of view is $\sim$11$'$ wide.
The image was obtained with the CTIO 1.5~m telescope through 
an $R_c$ filter, on 1999 Feb. 19.
\label{fi:std_sn99M}}
\end{figure}

\begin{figure}[t]
\plotone{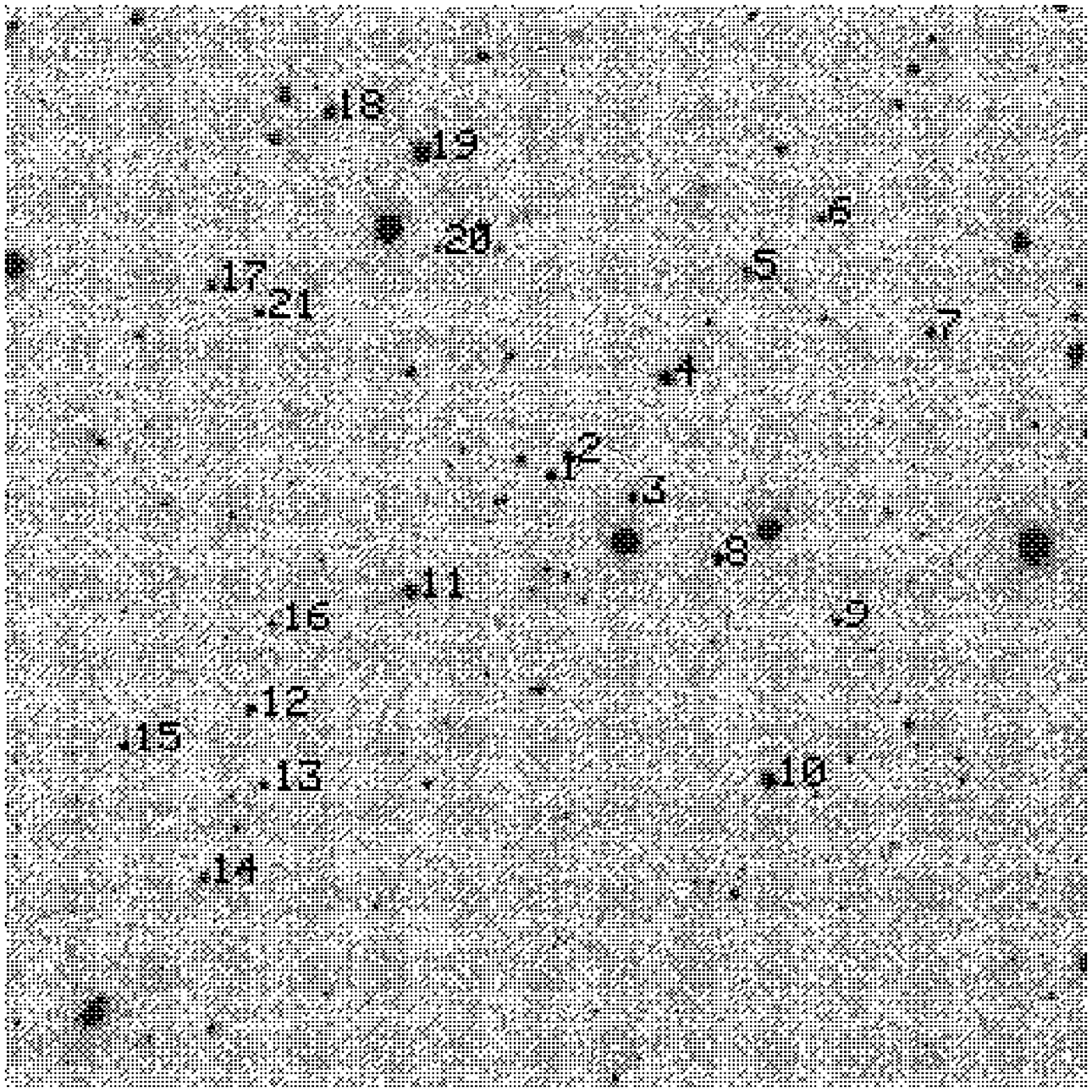}
\caption{Sequence of local standard stars in the field of SN~1999N.
North is up, east is to the left.
The field of view is $\sim$11$'$ wide.
The image was obtained with the CTIO 1.5~m telescope through 
an $R_c$ filter on 1999 Feb. 19.
\label{fi:std_sn99N}}
\end{figure}

\begin{figure}[t]
\plotone{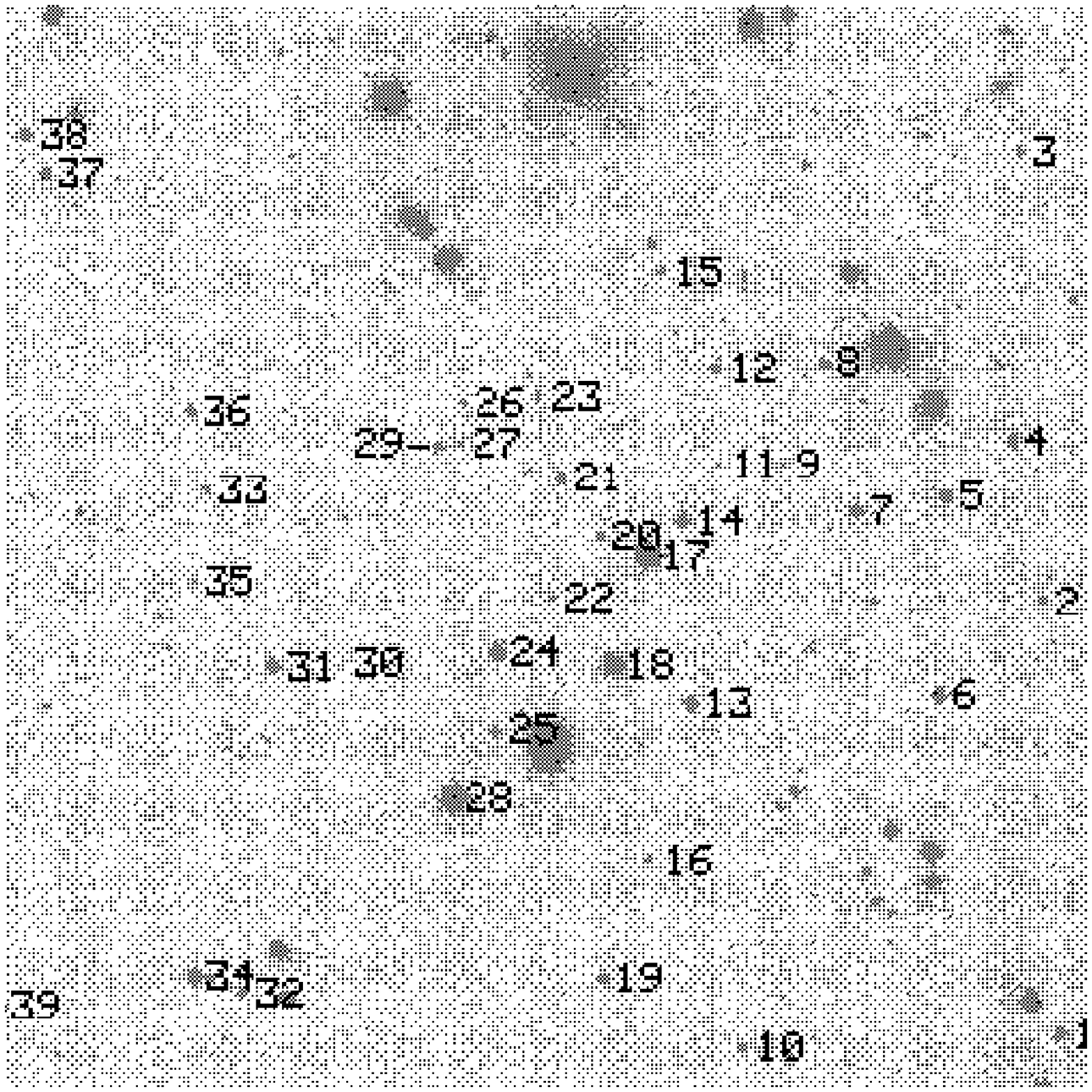}
\caption{Sequence of local standard stars in the field of SN~1999Q.
North is up, east is to the left.
The field of view is $\sim$11$'$ wide.
The image was obtained with the CTIO 1.5~m telescope through 
an $I_c$ filter on 1999 Feb. 18.
\label{fi:std_sn99Q}}
\end{figure}

\begin{figure}[t]
\plotone{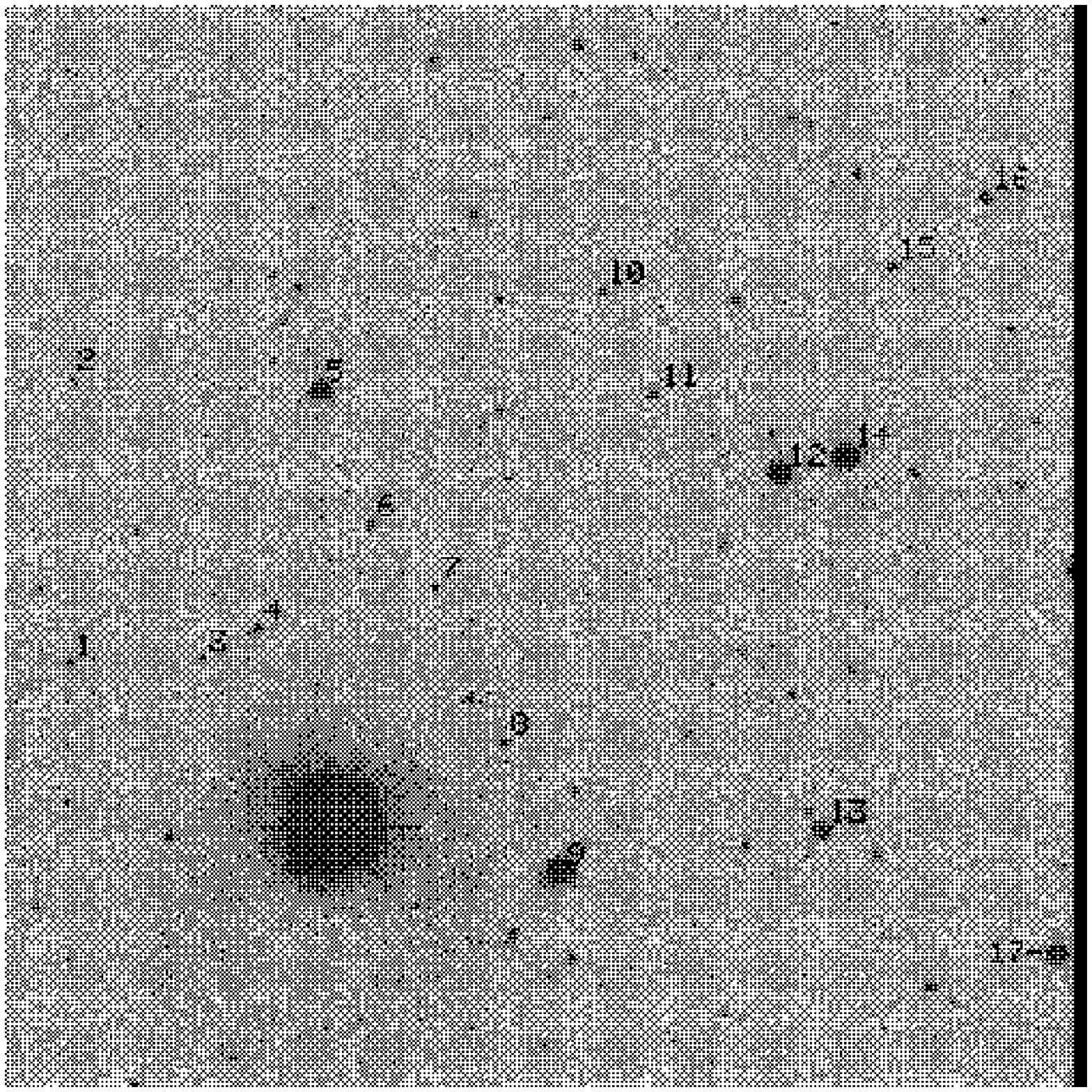}
\caption{Sequence of local standard stars in the field of SN~1999S.
North is up, east is to the left.
The field of view is $\sim$11$'$ wide.
The image was obtained with the CTIO 1.5~m telescope through 
an $R_c$ filter on 1999 Feb. 19.
\label{fi:std_sn99S}}
\end{figure}

\begin{figure}[t]
\plotone{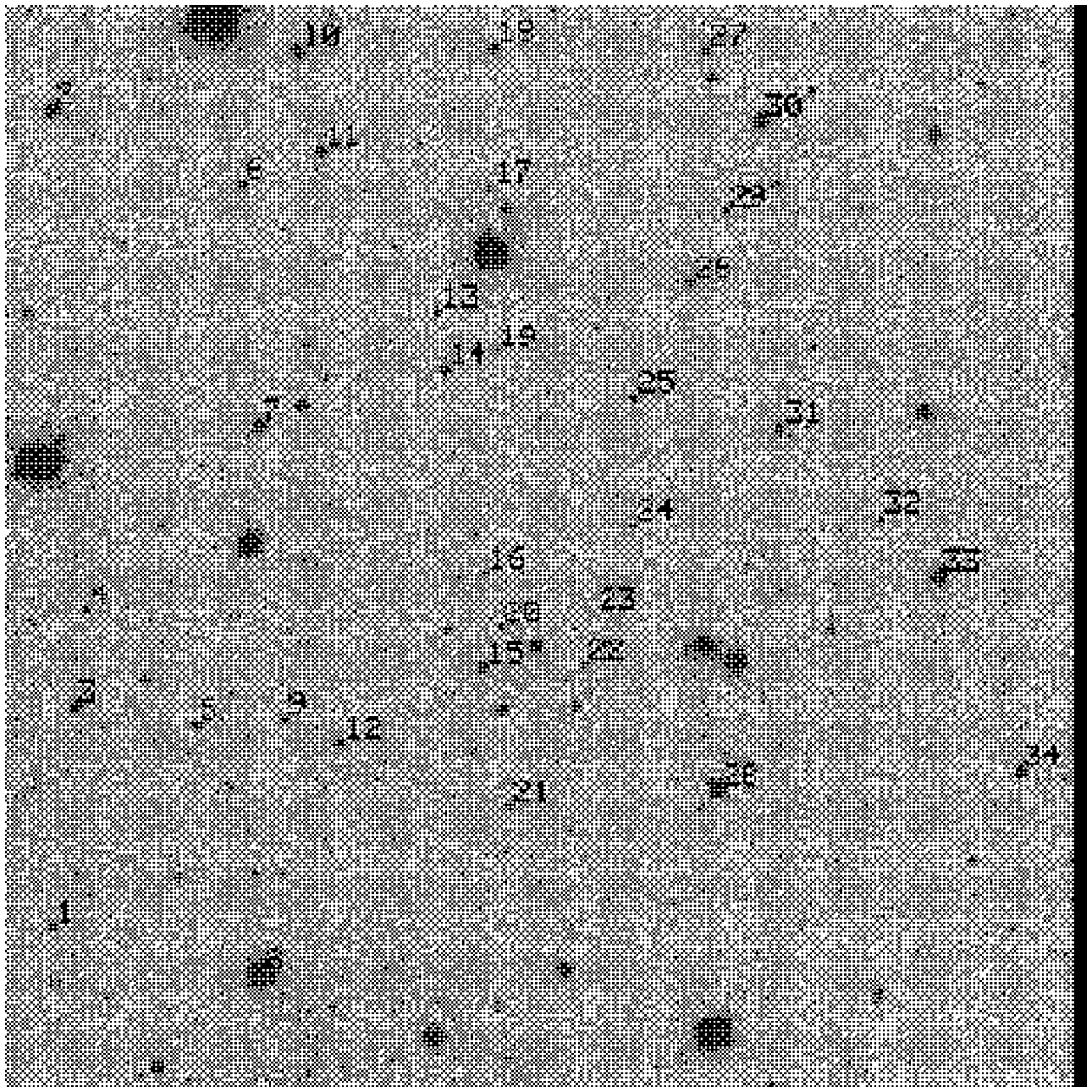}
\caption{Sequence of local standard stars in the field of SN~1999U.
North is up, east is to the left.
The field of view is $\sim$11$'$ wide.
The image was obtained with the CTIO 1.5~m telescope through 
an $R_c$ filter on 1999 Jan. 20.
\label{fi:std_sn99U}}
\end{figure}

For the local standard sequences of the SNe that were observed both
from ground-based telescopes and {\it HST}, therefore, we had
two independent measurements to check the calibration.
We compared the photometry of all stars in common between the
{\it HST} WFPC2 and ground-based fields,
finding the results to be consistent with the relations
provided by Holtzman et al. (1995) for ground-based and 
{\it HST} stellar photometry as a function of color index.
Suntzeff et al. (2006, in preparation) provide more detail of this analysis,
which increases the reliability of our photometric calibration.
Tables \ref{ta:std_sn99M} to \ref{ta:std_sn99U} present
$R$ and $R-I$ for the local sequences of standards,
and Figures \ref{fi:std_sn99M} to \ref{fi:std_sn99U} give their
finding charts.

\section{SN Photometry in the Observed Frame} \label{s:SN_phot}

\subsection{Images With and Without the Supernovae}

All the images of SN fields were
reduced following the standard techniques, with slightly
different procedures depending on the telescope used.
Observations of a given field were typically split 
into at least 3 images with small shifts
between them, to facilitate rejection of cosmic rays 
and CCD defects.

We had a special case where the number of
images to combine was unusually large.
Two runs with the CTIO 1.5~m telescope having nights
with a fairly bright moon were far from our other 
scheduled nights to do SN photometry, and were thus
important for the temporal coverage of the objects.
We obtained up to 30 images of each SN field,
with a total exposure time of $> 50$ min per field, to
greatly decrease the sky noise
(an approach resembling that used for infrared photometry).

Most of our ground-based
observations in the $I_c$ passband showed
substantial interference fringes.
In these cases, the observers used multiple dithered images in
different parts of the sky to prepare a ``fringe frame.''
This fringe frame was obtained by combining all the $I_c$
frames having the same exposure time on a given night
with a median rejection algorithm.
After subtraction of the fringe frame, the $I_c$ images displayed
a significantly flatter background.

Note that the noise pattern in the background
is altered when the fringe frame is subtracted.
In regions of the image were the fringes had been
determined by combining essentially empty sky, the 
root-mean-square (RMS) in the background is
unchanged.
In regions of the image where the fringe pattern emerged after
combining pixels with high signal (i.e., where bright stars or extended
objects were present), the RMS of the background usually increased.
We found increases of up to $\sim$15\% in the background RMS in
a few regions of some images.
These regions, however, were always far from the SNe,
since we had chosen to follow SNe that were far from bright stars or
bright extended objects.

The processed SN images were carefully analyzed by eye, with special
attention given to the regions of the SNe and host galaxies.
A few images were discarded because CCD or processing
defects were visible near the SN position.
In general, this did not significantly decrease the S/N; we
typically had many images
taken at similar epochs with different telescopes.

In order to perform unbiased photometry, a good model of the
brightness distribution underneath the SN, for each
particular image, is necessary.
We built these ``templates'' from images
of the field taken either the previous season, or
more than a year after SN discovery when the SN had
faded well below the detection limit.
It is imperative that the template image have high S/N,
excellent seeing, and good PSF shape.
Table \ref{ta:templates} gives a list of template
images obtained for each SN and photometric band.
In some cases we had more than one of these templates,
allowing us to study the effect that template matching and subtraction has
on the final photometry.

Template images taken with ground-based telescopes were reduced
and combined following the standard procedures.
The templates obtained with {\it HST} required special treatment.
Originally, one very late-time visit had been scheduled to take 
images of the field and host galaxy when the SN had faded away.
This last visit consisted, for each SN field and filter, of two images
with the same exposure parameters as had been used for the longest 
images taken when the SN was still bright.
Hence, we expected the same S/N in the template image as in the 
best images containing SNe.
One would actually prefer a template with significantly
{\it higher} S/N than the SN images, but this was a reasonable 
compromise given the limited amount of time allocated to our program.
Unexpectedly, however, the {\it HST} images taken in March 2000 
were noisier than the images of early 1999.
Subtraction of the templates would have produced a considerable 
reduction in the quality of our {\it HST} SN photometry.

Instead, we proceeded as follows.
We registered and combined all the {\it HST} images for each field, 
and used this very high S/N image to determine a good center for the SN.
We constructed the PSF for Chip 3 of each {\it HST} field (the chip containing
the SN) using all the isolated stars available on that chip.
We then fitted the PSF to the SN and subtracted it from the image, leaving 
the center fixed at the position computed in the combined image.
Next we combined all the images with the SN subtracted, and checked 
very carefully (line by line, and column by column) the place where 
the SN had been.
We found, in general, an increase of the noise at this position, but 
no systematic anomalies with respect to the pixels in regions not 
affected by the PSF subtraction.
When bad pixels or cosmic rays remained near the position of the SN, 
they were excised by hand.
After this, we combined all the images with the SN subtracted, 
together with those of March 2000, rejecting any remaining bad pixels 
and cosmic rays.
This combined image was our final template to subtract from the {\it HST} 
SN images.

To produce an {\it HST} template for ground-based images, we
convolved the {\it HST} template with a symmetric
kernel wider than the {\it HST} PSF, to make the resolution of 
the images $\sim$0.55$''$ (as explained below, this is required 
by the PSF-matching software to work).
These images are called ``version 2'' in Table \ref{ta:templates}.

\subsection{SN Photometry of {\it HST} Images}

Although noisy enough to prevent their adoption as templates, the 
March 2000 {\it HST} images of SN 1999Q and 1999U proved useful 
to simplify the SN photometry of the {\it HST} images.
We verified in the March 2000 images that no signal from the 
host galaxies was present within $0.5''$ of the SN position,
thus allowing us to use the typical strategies for {\it HST} 
stellar photometry.

We conducted PSF-fitting photometry with the package HSTPhot 
(Dolphin 2000), using the mid-2003 version of the package, 
zero points, and charge-transfer efficiency (CTE) correction 
parameters.
We applied HSTPhot to the images as they were provided by 
STScI, with the reduction pipeline updated as well to 
mid-2003 (Dolphin 2003).

As a check, we subtracted the {\it HST} template from the {\it HST} 
images and did aperture photometry of the SNe, applying the same 
zero points and CTE corrections as HSTPhot. We found, not surprisingly, 
that the results were consistent within the uncertainties.

Finally, the noisier March 2000 {\it HST} images were used to obtain
upper flux limits to the host-galaxy contribution,
by doing aperture photometry within
a radius of $0.5''$, the same as for the SN photometry.
We found the same upper limits for both SN~1999Q and SN~1999U:
$R = 26.19 \pm 0.50$ mag, and
$I = 25.76 \pm 0.35$ mag.

The {\it HST} photometry is included in Tables~\ref{ta:data99Q} and \ref{ta:data99U}.
Comparing the brightnesses of the SNe in the latest images with the upper limits
quoted above, we confirm that the SNe were at least approximately five times brighter than
the background.
In addition, we see that subtraction of the March 2000 images would have meant
uncertainties of $\sim$10\% in $R$ and $\sim$7\% in $I$, solely from the
template noise.
Not using template subtraction meant a modest improvement in the uncertainties, at
the expense of doing PSF-fitting photometry on a background essentially
that of the sky.

\subsection{Image Matching, Subtraction, and SN Photometry in Ground-Based Data}

For each SN passband set, one of the best images (SN well centered, good seeing)
was selected as astrometric reference, and all the other images were 
registered to match it.
The registration was done using IRAF tasks.
A registration map was built with GEOMAP
by using matched objects in image pairs, and then
the transformation was applied using GEOTRAN.
After registration, we compared the centers of the objects in both 
images and computed the RMS of the differences.
Our goal was to obtain an RMS smaller than 0.2 pixels.
For small fields of view, it was generally possible to meet this goal using
simple linear transformations (rotations plus translations).
For frames with a wide field of view, however, it was typically necessary to
fit distortion terms.

With the images registered, we matched the PSF and intensity scale of 
the template to those of the SN images.
For each image/template pair, we selected objects with good S/N
and compared their brightness distributions to
find the two-dimensional difference kernel that, when convolved 
with the image having the better seeing (usually the 
template), matches the PSF of the other.
We performed these tasks using a custom version of ISIS (Alard \& Lupton 1998).
ISIS builds a set of subrasters of the images around selected objects
(these are called ``stamps''), and compares the image/template pairs of 
stamps to find the difference kernel.
The version we used differs from the original in that it iterates the 
stamp-selection stage, checking the residuals of each stamp after 
matching is done, rejecting stamps that leave large residuals, and 
recomputing the kernel until convergence is obtained.
With the images matched in PSF shape and intensity, it was possible 
to subtract the template from the SN image and obtain, as a result, 
the SN PSF over an essentially constant background.

We found that ISIS performs well in situations where the full-width
at half maximum (FWHM) of the two compared PSFs is moderately different.
It does not provide good matches when they are too different, or when 
they are similar in FWHM but differ in other shape characteristics,
such as asymmetries oriented along different axes.
In particular, it was not possible to find a difference kernel to 
convert the {\it HST} PSF into a ground-based telescope PSF.
To obtain a successful matching between ground-based and space-based PSFs, 
it was necessary to degrade the resolution of the {\it HST} images, 
increasing the FWHM to $\sim$0.55$''$.
We did this by convolving the {\it HST} templates with a Gaussian 
symmetric kernel having a FWHM of 2 WFPC2 pixels.

Photometry of each SN was done on the subtracted image, but the image 
with the SN before background subtraction was used to fit the PSF 
shape and intensity scale.
For this, we followed the typical steps of PSF photometry, using the 
tasks of IRAF package DAOPHOT.
A set of stars suitable to do photometry was selected by eye,
centers for them were computed by fitting Gaussian kernels to 
the cores, and aperture photometry (without recentering) was done.
A subset of these stars was used to compute the PSF of the frame 
in the DAOPHOT style, with a theoretical function fitted to the 
core and a numerical residual computed to match the wings.
About 90\% of the time, the best-fitting theoretical function was 
a combination of a Gaussian core with Lorentzian wings (called 
``penny'' in IRAF tasks), and Moffat functions were the best fit 
in the remaining $\sim$10\% of the cases.
With the PSF computed, we performed PSF photometry
for all the selected stars in the image, and checked residuals 
to judge the quality of the PSF fitting.
The process of PSF fitting and subtraction was repeated until 
sufficiently small residuals were obtained.
In most cases, this required the use of a variable PSF,
depending on the angular size of the image and the combinations 
of telescope/instrument used in the template/image pair.
When computing the PSF, the star centers were left as a 
parameter to fit.

The PSF of the frame was then fitted to the residual image of the 
SN in background-subtracted frames.
The position of the SN was fixed, however, to avoid shifting of the 
PSF toward locally ``hot'' pixels (i.e., pixels that happened to
contain more signal due to statistical variations).
Uncorrected, this effect would produce systematically higher flux 
estimates and flatter light curves on the SN tails, when S/N is low.
The SN position was obtained by combining all the background-subtracted
images that showed a residual PSF after subtraction of the template.

We then generated a set of 20 artificial stars with the same flux as 
the SN, positioned them in places having background similar to that 
of the SN, and repeated the procedure to recover their instrumental 
magnitudes via PSF photometry.
This procedure allowed us to check the RMS in the SN photometry, and
to see whether a systematic intensity shift existed for the PSF of 
each particular frame.
With a couple of exceptions in the case of SN 1999M,
the shifts found were smaller than the RMS.

With the instrumental magnitudes of stars around each SN, we could 
fit a zero point and color term, using equations similar to 
Eq. (\ref{eq:calibration}).
When we had images of the SN in at least two passbands, we could invert 
the system and obtain the magnitude of the SN in both bandpasses.
In the few cases when we had only one passband for the SN, we interpolated
the other band from nearby epochs and used the interpolated value to compute
the color-term correction.
We used subsets of local standards with color similar to that of the SN to 
minimize the effect of the color terms and reduce the uncertainty.

\section{SN Photometry in the Rest Frame} \label{se:SN_rest}

\subsection{K-corrections and Light-Curve Fitting}

The procedure described above provided us with SN magnitudes in our frame of reference,
the observed frame.
To obtain the rest-frame magnitudes, which we need to compare with the local sample
of SNe~Ia, it is necessary to apply a K-correction.

We computed the K-corrections and fitted the SN light curves simultaneously.
We used two methods of light-curve fitting.
Each one of them employs a slightly different approach to the K-corrections.
In one case, cross-filter K-corrections as in Kim et al. (1996) and Schmidt et al. (1998)
are computed.
In the other, the method of Nugent et al. (2002) is closely followed.

As in Tonry et al. (2003), we write
the apparent brightness of a SN observed in filter $o$, $m_o(t)$, having a $z=0$
absolute magnitude light curve in filter $r$, $M_r(t^\prime))$, as

\begin{equation} \label{eq:Kcordef}
m_o(t) = M_r((1+z)t^\prime)+25 + 5\log\left( {d_{lum}} \over {\rm Mpc} \right) + 
K_{ro}((1+z)t^\prime),
\end{equation}
where $K_{ro}$ is given by

\begin{equation} \label{eq:newKcorr}
K_{ro} = 2.5\log\left[(1+z) {\int F_\lambda(\lambda) S_r(\lambda)d\lambda \over
\int F_\lambda(\lambda/(1+z))S_o(\lambda)d\lambda}\right] + {Z}_o - {Z}_r,
\end{equation}
for filter energy sensitivity functions $S_r$ and $S_o$, with zero
points ${Z}_r$ and ${Z}_o$, and SN spectrum $F_\lambda$.
The time scales $t$ and $t^\prime$ correspond to time after maximum brightness
in the observed frame and the SN rest frame, respectively.
We computed the $K_{ro}$ using sensitivity functions appropriate to each one of
the telescopes and instruments used, together with either a database of well
observed and timed spectra of nearby SNe~Ia, or the set of representative
SN~Ia spectra given by Nugent et al. (2002).

The SNe of this campaign were chosen so that their redshifts were close to
the redshift at which the observed-frame passbands $Rc$, $Ic$, $B45$, and $V45$,
are a good match to the rest-frame passbands $B$ and $V$.
This means that the integrals in Eq. (\ref{eq:newKcorr}) are approximately the 
same, and that their ratio is very close to one.
The dependence of the K-correction on the spectral energy distribution of 
the source is, hence, minimized.
This is important in two respects.
First, it minimizes the uncertainty in the K-correction
(Schmidt et al. 1998; Nugent et al. 2002).
Second, it minimizes any time dependence introduced
into the observed light curve by the K-correction.
The latter is necessary on general grounds, to avoid introducing the
time dependence of the nearby SNe~Ia (from which the K-corrections are
computed) into the distant SN~Ia light curves, but also because obtaining 
the K-correction for distant SNe~Ia is an iterative procedure.
The K-correction is a function of the intrinsic color of the SN, which is
in turn a function of the observed color of the SN and the K-correction. 
The intrinsic color of the SN depends on the foreground reddening and the
speed of the light curve, and both of them vary with the K-correction.
The fact that the rest-frame and observed-frame filters match well, 
however, makes the iteration of the K-corrections almost trivial.
The K-correction, the SN light curve,
the intrinsic color, and thus the foreground reddening
are simultaneously estimated.

Here we used two different light curve fitting methods:
the MLCS2k2 method, briefly described by Riess et al. (2004; see Jha, Riess,
\& Kirshner 2006 for details),
and the new PRES method, which is presented by Prieto et al. (2006).
The MLCS2k2 is the modern version of the MLCS method of Riess et al. (1998a, 2001).
Computation of MLCS2k2 distances was done together with K-corrections from
Nugent et al. (2002).
The PRES method is a generalization of the $\Delta m_{15}$ method of Hamuy et al. 
(1996a) and Phillips et al. (1999).
The main idea of the PRES method is to substitute the discrete set of SN light-curve 
templates by a continuous set of fitting templates built from the discrete
set by a weighted-average scheme.
The generalization allows a better fit to any observed light curve and, especially,
a more straightforward and elegant estimate of the fit uncertainties.
Computation of PRES distances was done together with K-corrections 
as in Kim et al. (1996) and Schmidt et al. (1998), but using a
more modern and extensive database with close to 150 spectra of nearby SNe~Ia
at different phases.
The uncertainties in the K-correction are computed from the scatter of
K-corrections for different SNe at the same phase.

The two light-curve fitting methods treat foreground reddening in different ways.
Before the light-curve fitting, both of the methods correct the observed color 
for Galactic reddening (Schlegel et al. 1998) as estimated in the direction of 
each SN.
The corrected observed color is the starting value of the iteration used to
estimate the host reddening.
The MLCS2k2 method uses a Galactic prior for the wavelength dependence
of the extinction law, an exponential prior on total extinction,
and a Gaussian prior on intrinsic color (see Riess et al. 2004, and
Jha et al. 2006 for further details).
SNe with heavy extinction are flagged
to avoid systematic errors from different types of dust.
In the PRESS method, the host-galaxy reddening is one of the parameters fitted
in a standard $\chi^2$ minimization.
A Galactic extinction law is used, but no prior on extinction or intrinsic
color is assumed.

\begin{figure}[th]
\plotone{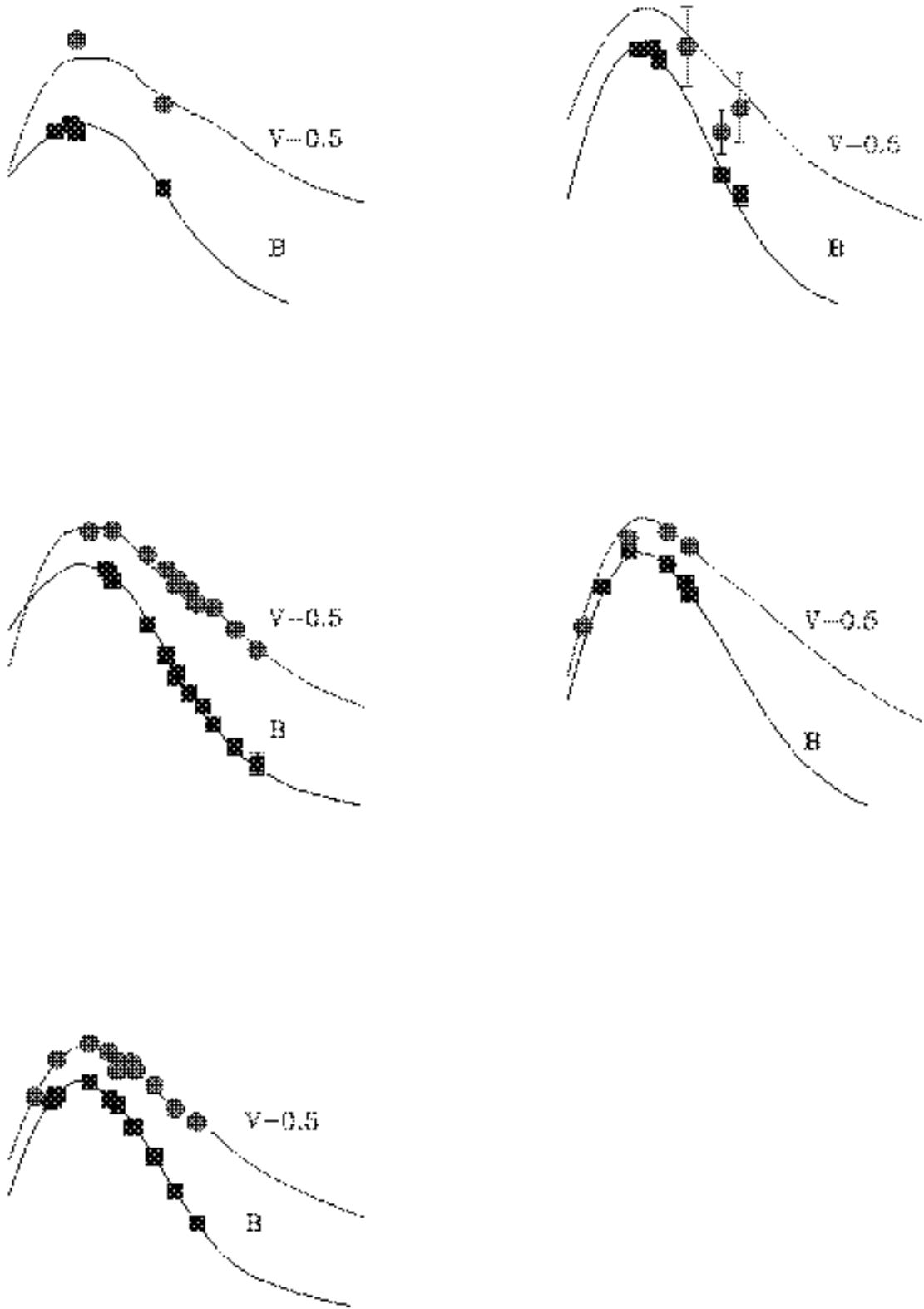}
\caption{SN light curves fitted using the PRES method.
The error bars correspond to $1\sigma$.
Where error bars are not visible, they are smaller than the
plotting symbol.
\label{fi:PRES_lightcurves}}
\end{figure}

The fact that the rest-frame and observed-frame filters are a good match 
implies that the dependence of our K-corrections on SN color was
very small.
This is good for both methods; it implies that the uncertainty was 
minimized.
Moreover, for MLCS2k2 there were generally only a few iterations 
between the light-curve fitting and K-correction routines.
The resulting SN photometry and K-corrections are given in Tables 
\ref{ta:data99M}--\ref{ta:data99U},
for distances calculated using both the PRES and the MLCS2k2 methods.
The light curves fitted using the PRES method are shown in Figure
\ref{fi:PRES_lightcurves}.
The light-curve fits using MLCS2k2 were similar; they
are not shown here.
Parameters fitted by the PRES method are given in Table~\ref{ta:table_pres},
and those fitted by MLCS2k2 in Table~\ref{ta:table_mlcs2k2}.

All of the SNe except SN~1999M provided reasonable light-curve fits.
An interesting cross-check is the phase that the epoch of maximum
light given by the fits implies for the spectra,
which can be compared with the phase of the best-matching spectra
given in Table~\ref{ta:spectra_data}.
The phases provided by the fits are within two days of those implied by
the spectra, the exception being again SN~1999M, which gives a mismatch
of $\sim 4.2$ days.

The mismatch of SN~1999M is puzzling.
Its light curve appears too broad for a SN~Ia, and its color is too red.
The PRES method provides a marginally tolerable fit, with
large uncertainties in the parameters, while
MLCS2k2 fails to provide an acceptable fit.
The identification of SN~1999M as a SN~Ia seems certain, however,
since the spectrum does show the \ion{Si}{2} $\lambda4130$ line.
The redshift is secure as well; it was measured through the 
\ion{O}{2} $\lambda3727$ emission line in the galaxy and shows consistency
with the observed shift of the SN features.

With the available data, little can be said of SN~1999M, other than it is
clearly an outlier in the known distribution of properties; there is no
reasonable match with the nearby SN sample used to calibrate the 
luminosity versus light-curve shape relations.
Given the S/N, the spectrum in Fig.~\ref{fi:spec_sn99M} seems
normal.
The \ion{Si}{2} line correlates well with temperature evolution 
in the ejecta of SNe~Ia; its strength in SN~1999M indicates that, 
close to maximum light, the photospheric temperature was typical.
The red color is, hence, correctly interpreted by the light-curve 
fitting methods as a result of large foreground reddening.
Consistent with the very broad light curve is an unusually
blue intrinsic color and a high luminosity at maximum.
Thus, this appears to have been a very luminous SN, but with 
large error bars in the calibrated photometry.
In comparison with the expectations for the low-mass $\Lambda$CDM 
cosmology, SN~1999M is the most luminous discovered so far by the HZT.

A possible explanation for the SN~1999M mismatch is a problem 
with one of the images used as templates,
in particular the rest-frame $V$ (observations in $I_c$).
SN1999M appeared in a bright host galaxy.
If the template subtraction in the observed-frame $I_c$ was
flawed, leaving the residual images of the SN with 
significant galaxy contamination, then the result would be a 
brighter and redder SN.
The best way to test this possibility is to obtain new,
high-quality templates and repeat the photometry from the start, 
a procedure beyond the scope of this work.

\subsection{Luminosity Distances} \label{se:distances}

With rest-frame, reddening-free light curves, the
luminosities of the distant SNe can be estimated.
The comparison of the observed flux,
$\cal{F}$, with the luminosity, $\cal{L}$,
provides us with luminosity distances,
\begin{equation} \label{eq:D_L}
D_L = \left(\frac{\cal{L}}{4 \pi \cal{F}}\right)^{\frac{1}{2}},
\end{equation}
and luminosity distance moduli $\mu = m - M$ related to $D_L$ through
the definition of magnitude.
The distance moduli estimated using the PRES method are included
in Table~\ref{ta:table_pres}, and those estimated using MLCS2k2 in
Table~\ref{ta:table_mlcs2k2}.

The values of $D_L$ and $\mu$ depend on the cosmological parameters,
as discussed extensively by Schmidt \etal\ (1998) for a $\Lambda$ 
cosmology and by Garnavich et al. (1998b) for a cosmology with a
more general equation of state for the ``dark energy.''
Comparison of the observed values of $\mu$ with those predicted by 
different cosmologies therefore allows us to set constraints on the 
cosmological parameters.
It is worth stressing that the experiment we describe,
the interpretation of the Hubble diagram at high redshift in terms
of the dynamical components of the Universe, depends on the
evolution of $D_L$ with redshift (i.e., on the ratio of 
$\cal{L}$ to $\cal{F}$ in Eq. (\ref{eq:D_L})), and not on the
actual value of $\cal{L}$.
Thus, our results are independent of the value of the Hubble constant
(see discussions in Schmidt et al. 1998; Garnavich et al. 1998b).

\section{Discussion} \label{se:discussion}

\begin{figure}[t]
\includegraphics[totalheight=7.5in, angle=0]{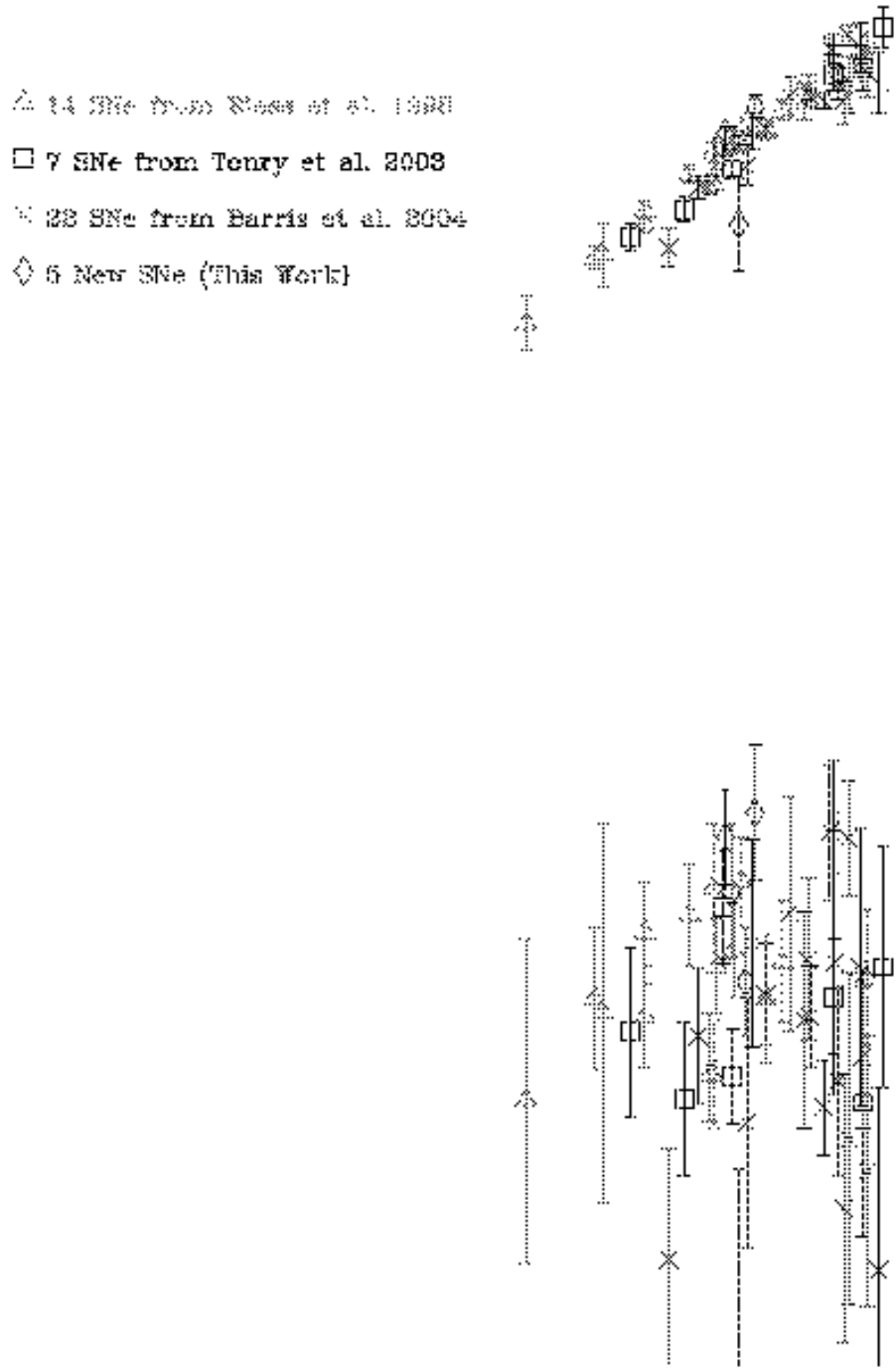}
\caption{Hubble diagram of the HZT SNe (this work, plus
Riess et al. 1998, Tonry et al. 2003, and Barris et al. 2004).
Error bars are $1\sigma$.
Models corresponding to the luminosity distances for a variety of 
parameters, as indicated, have been plotted.
The lower panel, with the same line style key, shows a differential
Hubble diagram, where the ordinate corresponds to the difference 
between a given distance modulus and that in an empty universe.
\label{fi:hubble_prs}}
\end{figure}

\begin{figure}[t]
\includegraphics[totalheight=7.5in, angle=0]{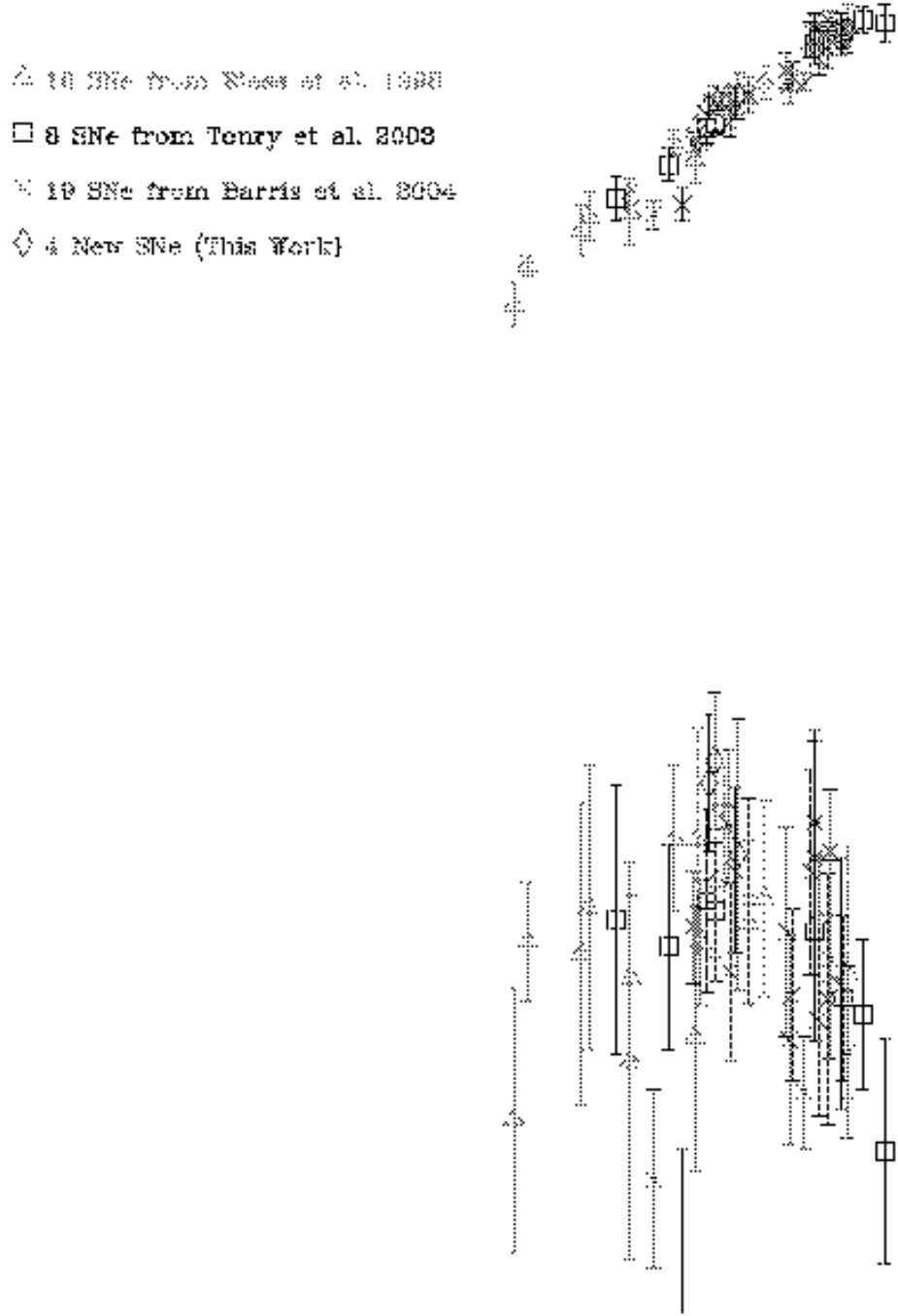}
\caption{The same as Figure \protect{\ref{fi:hubble_prs}}, but now 
for the distances calibrated using MLCS2k2.
\label{fi:hubble_mlcs2k2}}
\end{figure}

With the distance moduli computed, the SNe~Ia can be placed in the
high-redshift Hubble diagram that the HZT has been constructing 
since 1995.
In Figures \ref{fi:hubble_prs} and \ref{fi:hubble_mlcs2k2} we present the
Hubble diagram for distances determined with the PRES and MLCS2k2 methods,
respectively.
There is a different number of SNe in each plot, because the two methods 
use a different low-$z$ SN sample and, in addition, the highest-redshift 
bins require low-$z$ SNe with light curves calibrated in the ultraviolet,
a passband that has not been incorporated by the PRES method yet.
In both the MLCS2k2 and the PRESS methods used here, we have adopted 
$H_0 = 65$ km~s$^{-1}$~Mpc$^{-1}$.

All but one of the new SNe provide distances in excess of those expected 
in an empty universe, usually by a generous amount.
SN~1999M, the discrepant event, carries little weight in the Hubble diagram
because its distance modulus is either very uncertain (PRES light-curve 
calibration) or rejected from the light-curve fit (MLCS2k2 light-curve 
calibration).
The weighted average of the PRES luminosity distance moduli
for the five new SNe presented here is $42.64 \pm 0.08$ mag 
at an average redshift of $0.493 \pm 0.001$ ($1\sigma$ uncertainties).
This is 0.50 mag in excess of the distance modulus that a ``cold dark 
matter'' cosmology with $\Omega_M = 0.3$ (and $\Omega_\Lambda  = 0$) 
predicts, a result about 6$\sigma$ away from this particular expectation.
Allowing for the intrinsic dispersion of SN~Ia maximum brightness, the 
expected uncertainty of the five SNe increases to $\sim$0.1 ($1\sigma$),
which is still too small to account for the observed mean difference.  

One the one hand, our result is reassuring with respect to some 
concerns about systematics.
It is difficult to imagine {\it any} systematic effect related to 
photometry, calibration uncertainty, or evolutionary trends of SNe 
with lookback time that could give rise to such a large difference.
Thus, our results support the conclusion that the expansion of the
Universe is currently accelerating.

On the other hand, the average PRES distance modulus is too large to 
fit comfortably even within the $\Lambda$CDM concordance model.
If compared with the expectation for a $\Lambda$CDM cosmology,
with $\Omega_M = 0.3$ and $\Omega_\Lambda = 0.7$, the combined
distance modulus of the five SNe
is too high by 0.27 mag (i.e., about $3\sigma$ away).
%
%
The discrepancy remains when the calibration using the MLCS2k2 
method is considered.
The average distance modulus for the four SNe fitted by MLCS2k2 is
$42.74 \pm 0.11$ mag, which provides an excess of 0.58 mag with respect 
to the $\Omega_M = 0.3$ and $\Omega_\Lambda = 0.0$ model, a $6\sigma$ result.
If compared with the currently more realistic $\Lambda$CDM model 
($\Omega_M = 0.3$, $\Omega_\Lambda = 0.7$), the mean MLCS2k2 
luminosity distance is too high by 0.37 mag (nearly $4\sigma$).

Thus, the point in the Hubble diagram with high statistical significance 
at $z \approx 0.5$ that we sought to establish in this campaign turned
out to be challenging.
It not only appears inconsistent with a low-mass universe without 
a cosmological constant, but also with the currently favored 
concordance $\Lambda$CDM cosmology. 

The fact that this inference is based on only a handful of SNe does not reduce the 
impact of this particular campaign in the HZT Hubble diagram.
We may consider all the SNe with $0.44 < z < 0.54$ studied by our team to compute
a mean distance modulus at $z \approx 0.50$.
This redshift provides the best match between the rest-frame $B$ and 
observed-frame $R_c$ passbands, thus minimizing the dependence of the 
K-corrections with the supernova spectrum.
As a result, the K-correction is almost constant and its uncertainty is
minimized.
There are three SNe within this redshift range in Riess et al. (1998a), two in
Tonry et al. (2003), two in Barris et al. (2004), and four fitted by MLC2k2
in the present work.
The average distance modulus of these eleven SNe,
using the calibration provided by MLC2k2, is 
$42.51 \pm 0.07$ ($1\sigma$ uncertainty)
at a weighted mean redshift of 0.48.
%
%
This is, again, long by a generous margin:
close to $6\sigma$ away from the low-mass universe
($\Omega_M = 0.3$, $\Omega_\Lambda = 0.0$), 
and almost $3\sigma$ in excess of the concordance 
$\Lambda$CDM cosmology ($\Omega_M = 0.3$, $\Omega_\Lambda = 0.7$).
Considering only the SNe in Riess et al. (1998a), Tonry et al. 
(2003), and Barris et al. (2004), the mean distance modulus is
$42.39 \pm 0.08$ ($1\sigma$ uncertainty), for a 
weighted mean redshift of, again, 0.48.
The latter value is within $1\sigma$ of the expectation of the 
concordance model.

The combined distance moduli of all HZT SNe at $z \approx 0.48$,
with and without the SNe presented here, are consistent within 
$1\sigma$.
It is clear, however, that this particular campaign increases the 
shift to fainter magnitudes in the Hubble diagram at $z \approx 0.5$,
implying a stronger acceleration than that of the concordance model.

We fitted the data of figures \ref{fi:hubble_prs} and 
\ref{fi:hubble_mlcs2k2} with a grid of cosmological models 
varying $\Omega_M$, $\Omega_\Lambda$, and $H_0$.
We computed $\chi^2/N$ (where $N$ is the number of degrees of 
freedom), confirmed as in Tonry et al. (2003) that it is close 
to unity, then converted the $\chi^2$ distribution to a probability
distribution using 
${\rm exp}{(-\chi^2/2)}$, normalized the latter, marginalized 
over $H_0$, and found the contours of constant probability that
enclose 99.7\%, 95.4\%, and  68.3\% confidence.
The procedure is justified in detail by Riess et al. (1998a).

\begin{figure}[t]
\plotone{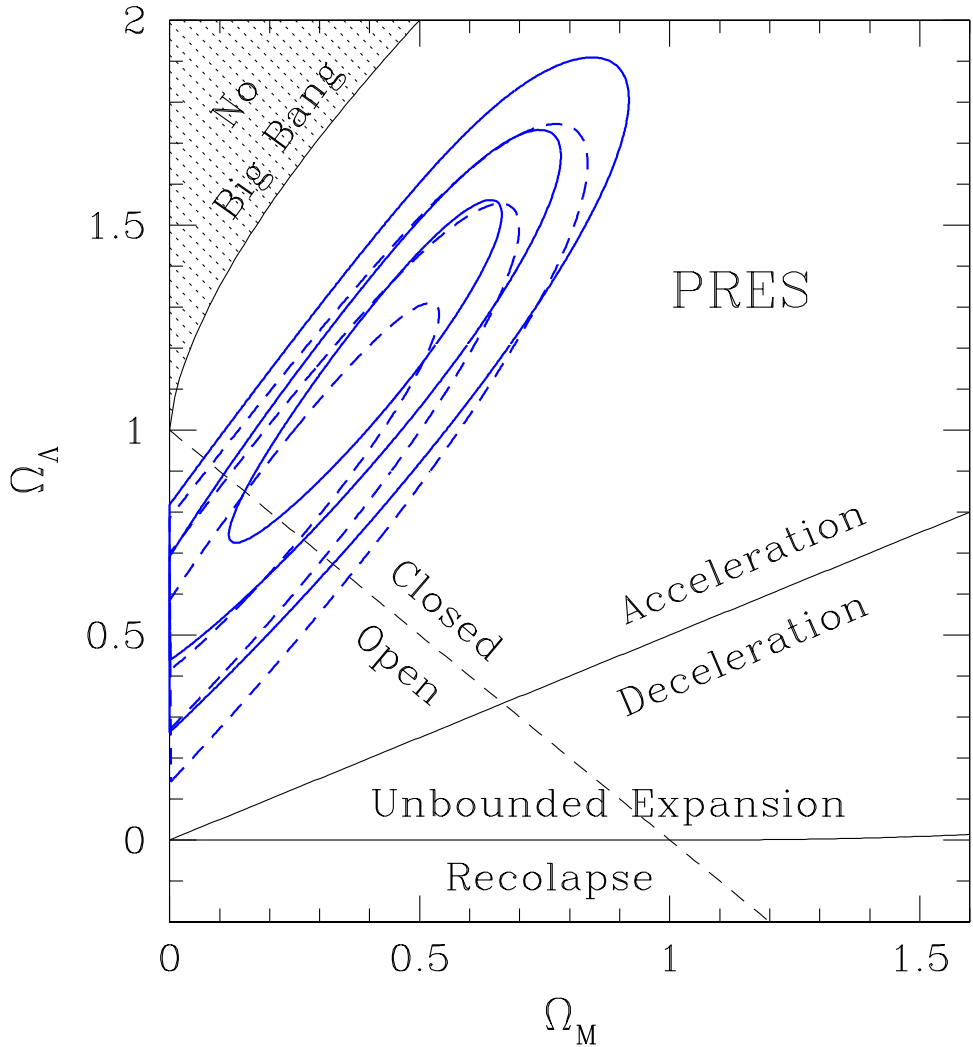}
\caption{Joint probability contours for the parameters
$\Omega_\Lambda$ and $\Omega_M$ that best fit the
Hubble diagram of Figure \protect{\ref{fi:hubble_prs}}
(distances calibrated using the PRES method).
>From larger to smaller, the contours drawn correspond
to 99.5\%, 97\%, and 68\% confidence, respectively.
Continuous lines show the contours corresponding to all
the HZT SNe for which a PRES light curve can be fitted.
Dashed lines correspond to the subset presented in
Riess et al. (1998), Tonry et al. (2003),
and Barris et al. (2004) (i.e., this campaign excluded).
The best-fitting parameters are given in Table~\protect{\ref{ta:cosmology}}.
\label{fi:contours_pres}}
\end{figure}

\begin{figure}[t]
\plotone{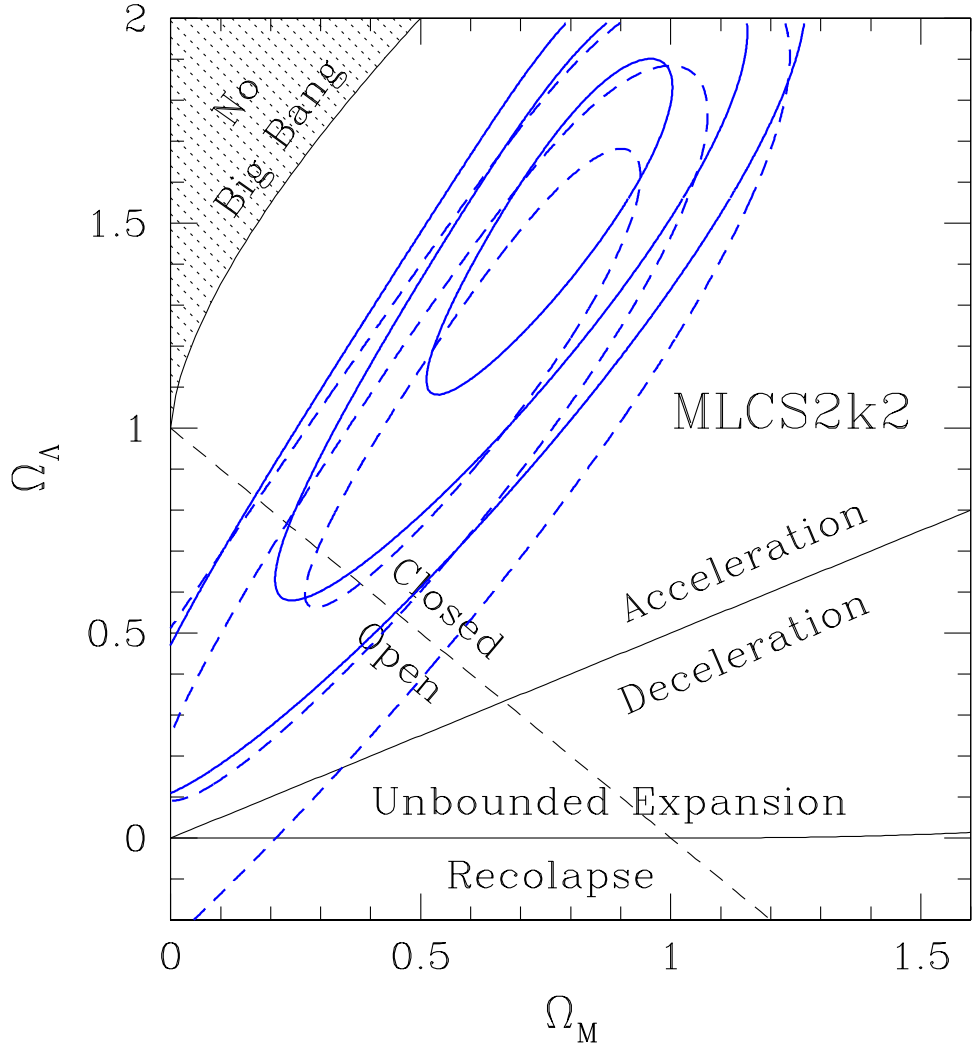}
\caption{The same as figure \protect{\ref{fi:contours_pres}},
but now the probability contours were fitted using the
Hubble diagram of Figure \protect{\ref{fi:hubble_mlcs2k2}}
(with distances calibrated using MLCS2k2).
The best-fitting parameters are given in Table~\protect{\ref{ta:cosmology}}.
\label{fi:contours_mlcs2k2}}
\end{figure}

We computed contours for the probability distributions implied by all
the HZT SNe~Ia in figures \ref{fi:hubble_prs} and \ref{fi:hubble_mlcs2k2},
for distances calibrated by both the PRES and the MLCS2k2 methods.
We present the results in Figures \ref{fi:contours_pres} and 
\ref{fi:contours_mlcs2k2}, for two different samples of the 
HZT SNe: (a) all of those in Riess et al. (1998a), Tonry et al. (2003), 
and Barris et al. (2004), and (b) these same SNe plus the new ones
reported in this paper.
The best-fit parameters for a $\Lambda$CDM cosmology resulting from the
fits are given in Table~\ref{ta:cosmology}, for both distance calibration
techniques and the different samples considered.
In this table, we provide the results of additional fits done with the
constraint that $\Omega_M$ + $\Omega_\Lambda = 1$, as suggested most
compellingly by the analysis of the WMAP observations (Spergel \etal\ 
2003).

The contours of figures \ref{fi:contours_pres} and \ref{fi:contours_mlcs2k2}
confirm the results obtained from inspection of the Hubble diagrams.
If the fits are left unconstrained, the HZT SNe tend to favor universes 
with high values of $\Omega_\Lambda$, but also high values of $\Omega_M$.
The new SNe in this campaign push the expected values of both $\Omega_\Lambda$
and $\Omega_M$ to somewhat higher values.

The different samples and fitting techniques also give slightly
different results.
Results from different samples are consistent within the $1\sigma$ 
uncertainty.
The different fitting techniques, however, lead to more significant 
differences, which are due in part to the fact that the SN samples 
differ in a sensitive way: the more distant SNe~Ia in Riess et al. 
(1998a), Tonry et al. (2003), and Barris et al. (2004) are included in 
the MLCS2k2 set, but not in the PRES set.
Finally, we note that the contours obtained with the MLCS2k2 distances
(figure \ref{fi:contours_mlcs2k2}) look very similar to the contours 
of Tonry et al. (2004) (see their Figure 10), even though the SN samples 
are different.

Rather than thinking in terms of the $\Omega$ parameters, which 
correspond to a definite model of the dynamical components of the 
Universe, we may simply measure its kinematics through the relation
between scale factor and time, represented by redshift and distance 
moduli, respectively.
As already mentioned by Tonry et al. (2003), we then conclude that 
our SNe suggest a universe that had both substantial deceleration at 
$z > 0.5$ and unambiguous acceleration more recently.
In particular, the SN data for $z \la 1.0$ provide strong
evidence that the expansion of the Universe was decelerating
at early times (Riess et al. 2004).
The low-redshift SNe ($z < 0.5$), of course, provide the best
evidence for recent acceleration, as was evident in both the
original HZT sample (Riess et al. 1998a) and the SCP sample 
(Perlmutter et al. 1999).
Our results presented here strengthen the case for recent 
acceleration.

The $\chi^2$ minimum in the elongated direction of the contour 
ellipses of figures \ref{fi:contours_pres} and 
\ref{fi:contours_mlcs2k2} is shallow.
Small-number statistics, as well as systematics present in a 
given redshift range or in a given search campaign, can easily 
move the best-fitting $\Omega$ values along this direction 
by a few tenths.
The clear shifts of the contours with changes in the sample
(in each figure, different line styles correspond to different 
samples) and with changes in the fitting technique (compare
the various figures) illustrate the sensitivity of the results 
to these issues.

The discrepancy between the results of MLCS2k2 and PRES
especially concerns us.
To clarify this point, we repeated the computation of contour 
levels for a sample consisting of exactly the same SN set 
(92 SNe, 48 of which are in the low-$z$ sample).
We do not show the results here because they are redundant:
the contours are almost the same as those shown in figures
\ref{fi:contours_pres} and \ref{fi:contours_mlcs2k2}.
Thus, it appears that the discrepancy is really due to the
different methods used.
Further work, already in progress, will help clarify this issue.

The SN sample analyzed here provides an excess distance with respect 
to the concordance $\Lambda$CDM cosmology as large as that given by
the sample of Riess et al (1998a) with respect to the empty Universe.
Should this be a cause of concern? We think not, mainly for two reasons.

First, the strength of the concordance model comes largely from
the convergence of various independent experiments that provide 
consistent expectation values for the cosmological parameters, or for
combinations of them.
Measuring luminosity distances versus redshift provides a good 
constraint for combinations of the form $\Omega_\Lambda - \Omega_M$ 
(Schmidt et al. 1998).
It is reassuring to see that, regardless of the chosen sample,
the elongated axis of the contour ellipses always crosses the 
line $\Omega_\Lambda + \Omega_M = 1$ at values consistent with 
$0.2 < \Omega_M < 0.3$, right in the interval favored by several
experiments measuring $\Omega_M$.

Second, the size of the contours and the position of the 
probability maximum strongly depend both on the number and
quality of the SNe~Ia in the sample.
When the HZT SNe are augmented with those discovered and studied 
by other programs (see Table 5 in Riess et al. 2004, and references 
therein), the probability contours shrink, the best-fitting values of 
both $\Omega_\Lambda$ and $\Omega_M$ decrease, and the plot approaches 
that expected for the concordance model.
The presence of several SNe at $z > 1$ is especially important in 
producing this effect.

\begin{figure}[t]
\plotone{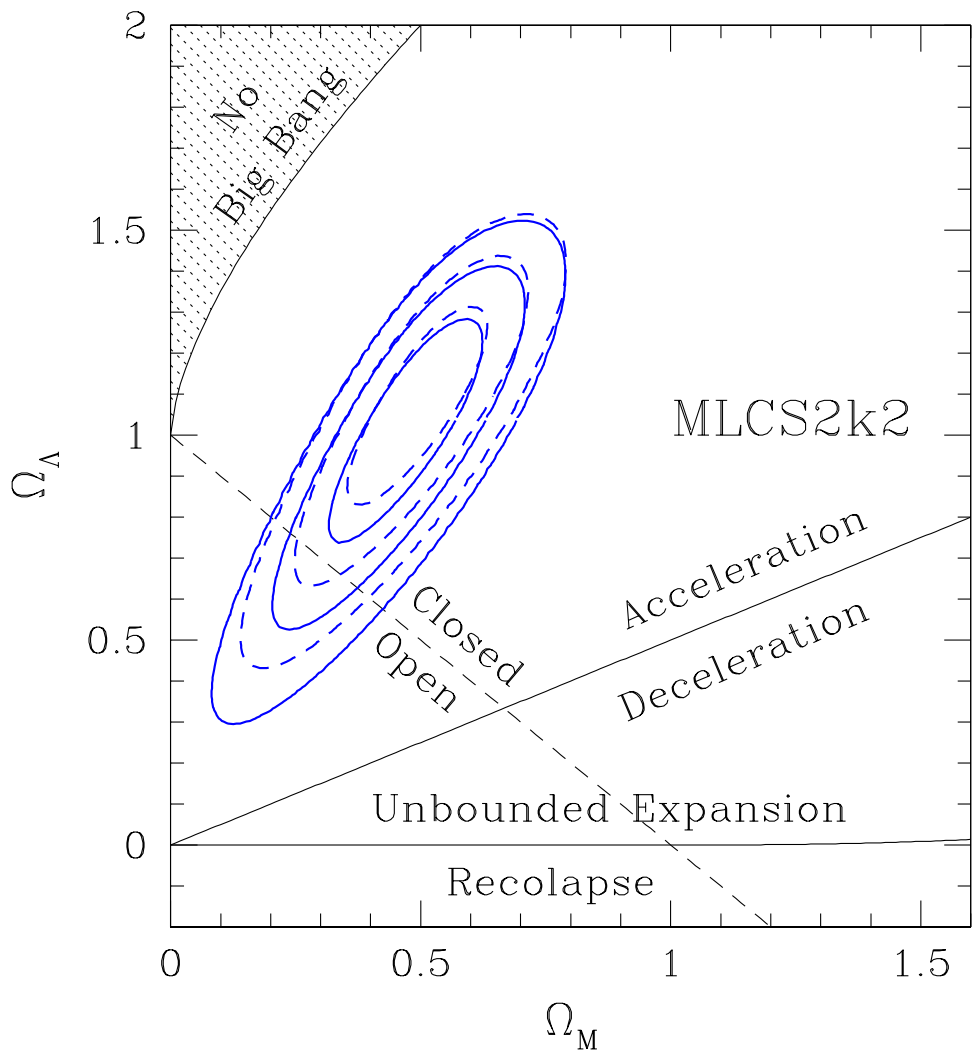}
\caption{The dashed line shows the joint probability contours for the
whole sample of ``gold'' and ``silver'' SNe of Riess et al. (2004), plus
SNe~1999N, 1999Q, 1999S, and 1999U.
The solid line shows only the ``gold'' SNe of Riess et al. (2004), plus
the same four SNe of this campaign.
In both cases, distances were calibrated using MLCS2k2.
\label{fi:contours_mlcs2k2_all}}
\end{figure}

Figure \ref{fi:contours_mlcs2k2_all} shows the results of testing
this second assertion.
As the number of high-quality SNe using the ``gold'' sample of 
Riess et al. (2004) increases, the area enclosed by the confidence 
contours decreases.
In particular, the major axis becomes shorter faster than the
minor axis, approaching the minor axis in length; the minimum along
the major axis is made deeper by increasing the sample.
Moreover, there is a shift of the best-fitting $\Omega$ values 
toward the line $\Omega_\Lambda + \Omega_M = 1$.
This suggests that the high values of $\Omega_\Lambda$ and 
$\Omega_M$ implied by the HZT Hubble diagrams of high-$z$ SNe~Ia 
need not be taken too seriously; they are still quite dependent on 
the sample size, and especially on the number of SNe in the 
highest-redshift bins.

\begin{figure}[t]
\includegraphics[totalheight=7.5in, angle=0]{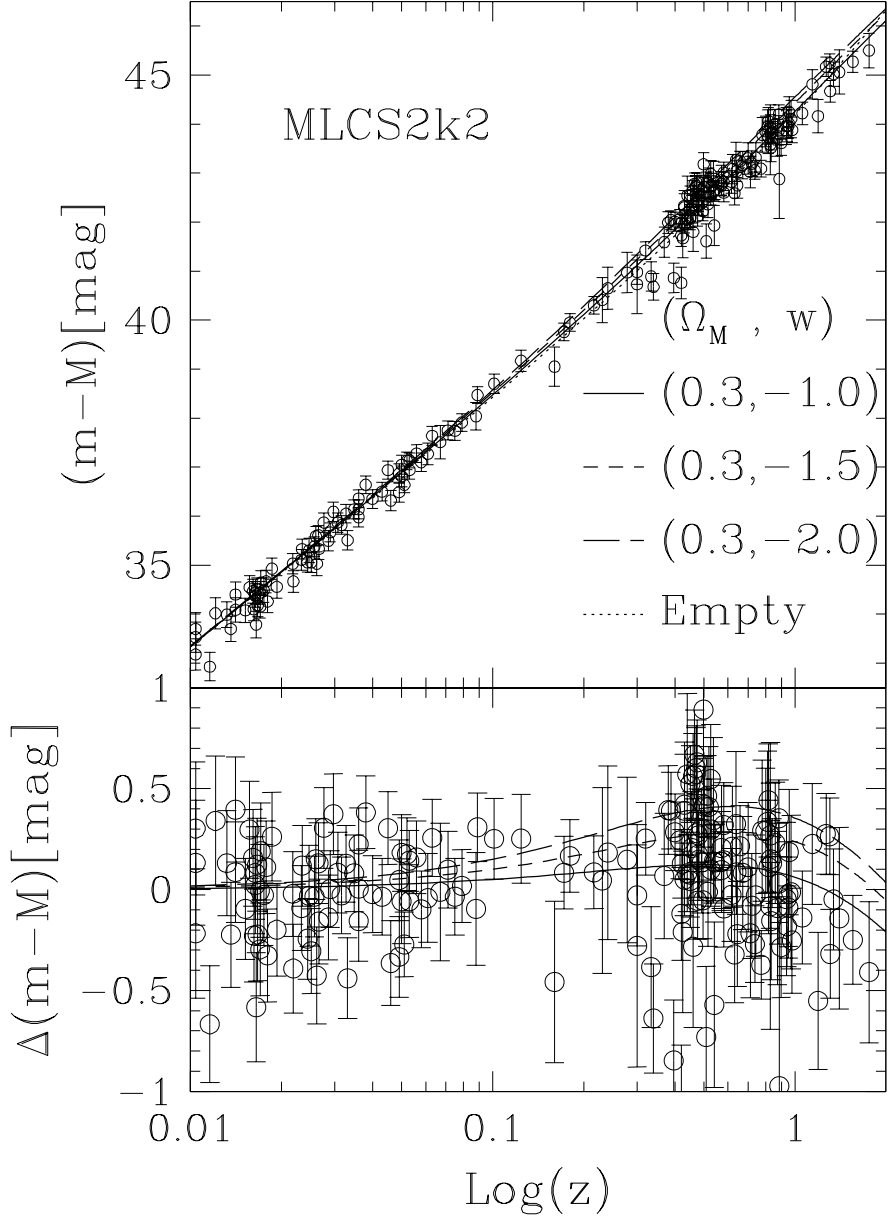}
\caption{The Hubble diagram including the SNe presented in this study, 
using solid symbols, and those of Riess et al. (1998) and Tonry et al. 
(2003), using open symbols.
Error bars correspond to $1\sigma$.
The upper panel shows the standard diagram, including the curves 
corresponding to the luminosity distances for universes with various 
parameters, as given in the legend.
The lower panel shows the differential Hubble diagram, where the 
ordinate corresponds to the difference between the distance moduli 
observed (for SNe) and computed (for models) with the empty universe.
The legend for line types is the same as in the upper panel.
\label{fi:hubble_w}}
\end{figure}

Even within the limits of the sample, and bearing in mind the
differences implied by different fitting techniques, one may
ask whether the failure of the Hubble diagrams of various
SN subsets to provide best-fitting values of $\Omega_M$ and 
$\Omega_\Lambda$ closer to those expected from the concordance 
model could be a sign of weakness in the new $\Lambda$CDM paradigm.
Specifically, perhaps the dark energy is not the cosmological
constant, but rather some kind of ``quintessence'' or rolling
scaler field.
We explored this question following Garnavich et al. (1998b),
considering universes with only gravitating matter ($\Omega_M$)
and a dark-energy component ($\Omega_x$). 
We assumed further that the Universe is flat, and that
$\Omega_x$ can be approximated with an equation of state of 
the form $P_x = w_x \rho_x c^2$, where $P_x$ and $\rho_x$ are 
the pressure and density of the dark-energy component.
(Note that $w = -1$ if the dark energy is indeed the 
cosmological constant.)
With these elements a new set of models for the luminosity 
distances can be computed and the predictions compared with 
the observations.
In Figure \ref{fi:hubble_w} we show the results of this 
exercise.

It is not surprising that abandoning the assumption that 
$w_x = -1$ results in a better fit of the offset of SN distances 
at $z \approx 0.5$, since there is now a new parameter to adjust.
The interesting fact is that, if we want to fit the SN distances 
at $z \approx 0.5$, then we need to {\it decrease} $w_x$ to values 
significantly smaller than $-1$.
The best fit gives $w \approx -1.8$, well below that of a 
cosmological constant.
However, matching the luminosity distances at $z \approx 0.5$ 
with $w_x \approx -1.8$ leads to a bad fit at higher redshifts.

We see that allowing $w_x$ to be a free parameter does not 
provide the flexibility needed to match the apparently sharp 
transition in the behavior of luminosity distances with redshift 
at $z \approx 0.5$.
But at this point, our result is only illustrative at best;
the samples considered here are not sufficiently large, 
homogeneous, and well-distributed in redshift to accurately 
measure $w_x$.
The point of figure \ref{fi:hubble_w} is to show that in a 
low-mass, flat universe, any static value of $w_x$ will fare 
as badly as $w = -1$ (that is, $\Lambda$) in matching the 
current Hubble diagram of high-$z$ SNe~Ia, especially the 
transition in luminosity distances over the redshift range 
0.5--1.0.

In summary, the new SNe~Ia presented in this paper support the 
previous conclusion of the HZT and the SCP that the expansion 
of the Universe is currently accelerating, consistent with the
presence of a cosmological constant or some other form of ``dark 
energy.''
It is encouraging to see that, as the number of SNe increases,
the contours of the HZT Hubble diagram, as well as those of 
other collaborations, become tighter and more consistent with 
the ``concordance Universe.''
However, to make further progress, we need more SNe, 
better light curves, more reliable spectroscopic 
classifications, and more accurate K-corrections.
As the samples grow, decreasing scatter caused by the intrinsic 
dispersion in luminosity and by observational uncertainties,
proper attention to systematic errors will become paramount.
In particular, it will become pressing to obtain a better 
understanding of the entire SN~Ia phenomenon.
This understanding will be needed on many fronts to
test the empirical calibrations that we so confidently 
extrapolate at high redshifts: from theoretical studies 
of the progenitors, their evolutionary path to the explosive
state, and the explosion itself, to detailed analytical studies 
of nearby and progressively more distant SN samples.

\acknowledgments

We thank the staffs of the many facilities we used
for their assistance with the observations.
Financial support for this work was provided by
NASA through grants GO-08177, GO-08641, and GO-09118 from the Space
Telescope Science Institute, which is operated by the Association of
Universities for Research in Astronomy, Inc., under NASA contract NAS
5-26555.
Funding was also provided by National Science Foundation
grants AST-0206329, AST-0307894, and AST-0443378.
A.C. acknowledges the support of
CONICYT (Chile) through FONDECYT grants 1980803, 1000524, and 7000524,
as well as the Directors of CTIO and MSSSO for support during periods of work in
La Serena and Canberra, respectively.
A.V.F. is grateful for a Miller Research Professorship at UC Berkeley 
during which part of this work was completed.

\newpage

\clearpage

\clearpage

\clearpage

\clearpage

\clearpage

\clearpage

\clearpage

\clearpage

\clearpage

\clearpage

\clearpage

\clearpage

\clearpage

\clearpage

\clearpage

\clearpage

\clearpage

\clearpage

\clearpage

\clearpage

\clearpage

\clearpage

\begin{deluxetable}{lrrcccc}
\tablewidth{0pt}
\tablecaption{Observations of the Early 1999 Campaign\label{ta:candidates}}
\tablehead{
\colhead{SN name} &
\colhead{RA (2000.0)} &
\colhead{Dec (2000.0)} &
\colhead{Discovery\tablenotemark{a}} &
\colhead{$E(B{-}V)$ (mag)} &
\colhead{$N_{obs}$\tablenotemark{b}} &
\colhead{HST}
}
\startdata
SN 1999M & 10 05 46.60 & $-$07 27 31.3  & 192.4 & 0.025  & 29 & no \\
SN 1999N & 10 04 27.93 & $-$07 35 51.1  & 192.4 & 0.028  & 24 & no \\
SN 1999Q & 08 00 45.31 & +05 31 27.4  & 193.2 & 0.021  & 83 & yes \\
SN 1999S & 09 28 08.08 & $-$04 58 37.6  & 193.3 & 0.035  & 53 & no \\
SN 1999U & 09 26 42.96 & $-$05 37 57.8  & 193.2 & 0.038  & 98 & yes \\
\enddata
\tablenotetext{a}{Mean Julian Date minus 2451000 days.}
\tablenotetext{b}{Number of individual images taken. Many were later combined.}
\end{deluxetable}

\begin{deluxetable}{lllrrccc}
\tablewidth{0pt}
\tablecaption{Spectroscopy of the Selected Type Ia Supernovae\label{ta:spectra_data}}
\tablehead{
\colhead{Name} &
\colhead{Redshift\tablenotemark{a}} &
\colhead{Redshift Method} &
\colhead{Exp. (s)\tablenotemark{b}} &
\colhead{Date\tablenotemark{c}} &
\multicolumn{3}{c}{Phase\tablenotemark{d}} \\
\cline{6-8} \\
\colhead{} & \colhead{} & \colhead{} & \colhead{} & \colhead{} &
\colhead{Sp.\tablenotemark{e} } &
\colhead{PRES\tablenotemark{f} } &
\colhead{MLCS2k2\tablenotemark{g}}
}
\startdata
1999M & 0.493(1) & [O~II] $\lambda3727$ in host  &  4800&198.0&\phs3& $-$1.3&\phs\nodata\\
1999N & 0.537(5) & SN features                               &  3000&198.0&\phs3&\phs3.8&\phs3.6\\
1999Q & 0.459(8) & SN features                               & 24100&198.5&\phs3&\phs6.1&\phs4.2\\
1999S & 0.474(1) & [O~II] $\lambda3727$ in host  &  7200&197.0& $-7$& $-$7.9& $-$8.5\\
1999U & 0.511(1) & [O~II] $\lambda3727$ in host  & 11400&197.5& $-7$& $-$6.2& $-$5.1\\
\enddata
\tablenotetext{a}{Uncertainty in the last digit given in parentheses.}
\tablenotetext{b}{Total exposure time in seconds.}
\tablenotetext{c}{Mean Julian Date of exposures minus 2451000 days.}
\tablenotetext{d}{Phase in rest-frame days after $B$ maximum.}
\tablenotetext{e}{Phase from spectroscopic match. Uncertainty estimated to be 4 days.}
\tablenotetext{f}{Phase implied by PRES light-curve fitting (see \S \protect{\ref{se:SN_rest}}).
Uncertainty estimated to be 2 days.}
\tablenotetext{g}{Phase implied by MLCS2k2 light-curve fitting (see \S \protect{\ref{se:SN_rest}}).
Uncertainty estimated to be 2 days.}
\end{deluxetable}

\begin{deluxetable}{rccc}
\tablewidth{0pt}   
\tablecaption{ Photometric Sequence Near SN~1999M \label{ta:std_sn99M}}
\tablehead{
\colhead{ID} &
\colhead{$R$}  &
\colhead{$R-I$}} \startdata
 1 & 14.808 $\pm$ 0.003 & 0.372 $\pm$ 0.004\\
 2 & 16.360 $\pm$ 0.003 & 1.132 $\pm$ 0.004\\
 3 & 18.078 $\pm$ 0.007 & 0.916 $\pm$ 0.009\\
 4 & 18.661 $\pm$ 0.017 & 0.483 $\pm$ 0.020\\
 5 & 18.169 $\pm$ 0.006 & 0.802 $\pm$ 0.009\\
 6 & 18.285 $\pm$ 0.010 & 0.437 $\pm$ 0.011\\
 7 & 17.799 $\pm$ 0.003 & 0.517 $\pm$ 0.004\\
 8 & 19.152 $\pm$ 0.014 & 0.428 $\pm$ 0.039\\
 9 & 17.625 $\pm$ 0.003 & 1.457 $\pm$ 0.005\\
10 & 16.876 $\pm$ 0.004 & 0.564 $\pm$ 0.007\\
11 & 18.734 $\pm$ 0.004 & 0.952 $\pm$ 0.012\\
12 & 19.912 $\pm$ 0.006 & 0.765 $\pm$ 0.020\\
13 & 20.211 $\pm$ 0.030 & 1.127 $\pm$ 0.035\\
14 & 18.699 $\pm$ 0.007 & 0.894 $\pm$ 0.008\\
15 & 19.276 $\pm$ 0.010 & 1.102 $\pm$ 0.011\\
16 & 18.476 $\pm$ 0.011 & 0.766 $\pm$ 0.012\\
17 & 17.571 $\pm$ 0.007 & 1.183 $\pm$ 0.008\\
18 & 19.260 $\pm$ 0.020 & 0.412 $\pm$ 0.024\\
19 & 18.743 $\pm$ 0.004 & 0.362 $\pm$ 0.012\\
20 & 18.182 $\pm$ 0.007 & 0.985 $\pm$ 0.008\\
21 & 20.257 $\pm$ 0.007 & 0.749 $\pm$ 0.013\\
22 & 20.378 $\pm$ 0.015 & 0.678 $\pm$ 0.043\\
\enddata
\end{deluxetable}

\begin{deluxetable}{rccc}
\tablewidth{0pt}   
\tablecaption{Photometric Sequence Near SN~1999N \label{ta:std_sn99N}}
\tablehead{
\colhead{ID} &
\colhead{$R$}  &
\colhead{$R-I$}} \startdata
1 & 17.505 $\pm$ 0.01 & 0.401 $\pm$ 0.012 \\
2 & 17.73 $\pm$ 0.004 & 0.871 $\pm$ 0.007 \\
3 & 17.388 $\pm$ 0.004 & 0.798 $\pm$ 0.007 \\
4 & 16.245 $\pm$ 0.004 & 0.479 $\pm$ 0.007 \\
5 & 17.377 $\pm$ 0.004 & 0.582 $\pm$ 0.007 \\
6 & 17.571 $\pm$ 0.004 & 0.355 $\pm$ 0.008 \\
7 & 17.956 $\pm$ 0.005 & 0.408 $\pm$ 0.008 \\
8 & 17.106 $\pm$ 0.005 & 0.461 $\pm$ 0.007 \\
9 & 18.396 $\pm$ 0.006 & 0.496 $\pm$ 0.01 \\
10 & 16.268 $\pm$ 0.009 & 0.345 $\pm$ 0.01 \\
11 & 16.138 $\pm$ 0.004 & 0.368 $\pm$ 0.006 \\
12 & 17.914 $\pm$ 0.01 & 0.46 $\pm$ 0.012 \\
13 & 18.618 $\pm$ 0.004 & 1.632 $\pm$ 0.007 \\
14 & 17.434 $\pm$ 0.004 & 1.319 $\pm$ 0.007 \\
15 & 18.08 $\pm$ 0.007 & 0.878 $\pm$ 0.009 \\
16 & 18.836 $\pm$ 0.011 & 0.504 $\pm$ 0.013 \\
17 & 17.794 $\pm$ 0.005 & 0.381 $\pm$ 0.008 \\
18 & 16.954 $\pm$ 0.004 & 0.341 $\pm$ 0.007 \\
19 & 14.997 $\pm$ 0.004 & 0.396 $\pm$ 0.006 \\
20 & 18.898 $\pm$ 0.011 & 0.289 $\pm$ 0.015 \\
21 & 18.513 $\pm$ 0.004 & 0.931 $\pm$ 0.008 \\
\enddata
\end{deluxetable}

\begin{deluxetable}{rccc}
\tablewidth{0pt}   
\tablecaption{Photometric Sequence Near SN~1999Q \label{ta:std_sn99Q}}
\tablehead{
\colhead{ID} &
\colhead{$R$}  &
\colhead{$R-I$}} \startdata
1 & 18.947 $\pm$ 0.008 & 1.003 $\pm$ 0.011\\
2 & 19.983 $\pm$ 0.013 & 0.887 $\pm$ 0.020\\
3 & 20.105 $\pm$ 0.014 & 0.756 $\pm$ 0.020\\
4 & 18.925 $\pm$ 0.007 & 0.737 $\pm$ 0.011\\
5 & 18.933 $\pm$ 0.008 & 1.437 $\pm$ 0.021\\
6 & 19.016 $\pm$ 0.008 & 1.380 $\pm$ 0.011\\
7 & 19.186 $\pm$ 0.008 & 0.987 $\pm$ 0.012\\
8 & 19.889 $\pm$ 0.013 & 0.337 $\pm$ 0.023\\
9 & 21.471 $\pm$ 0.042 & 0.664 $\pm$ 0.065\\
10 & 19.753 $\pm$ 0.011 & 0.889 $\pm$ 0.017\\
11 & 21.210 $\pm$ 0.029 & 0.300 $\pm$ 0.063\\
12 & 19.875 $\pm$ 0.012 & 0.832 $\pm$ 0.018\\
13 & 18.471 $\pm$ 0.007 & 0.956 $\pm$ 0.010\\
14 & 18.218 $\pm$ 0.008 & 0.316 $\pm$ 0.014\\
15 & 21.174 $\pm$ 0.031 & 1.354 $\pm$ 0.037\\
16 & 20.792 $\pm$ 0.022 & 1.142 $\pm$ 0.031\\
17 & 15.355 $\pm$ 0.007 & 0.338 $\pm$ 0.010\\
18 & 17.216 $\pm$ 0.007 & 1.449 $\pm$ 0.010\\
19 & 19.867 $\pm$ 0.010 & 1.348 $\pm$ 0.014\\
20 & 21.313 $\pm$ 0.032 & 1.274 $\pm$ 0.042\\
21 & 19.639 $\pm$ 0.011 & 1.375 $\pm$ 0.014\\
22 & 21.198 $\pm$ 0.029 & 0.378 $\pm$ 0.069\\
23 & 21.699 $\pm$ 0.044 & 1.410 $\pm$ 0.057\\
24 & 17.103 $\pm$ 0.008 & 0.348 $\pm$ 0.011\\
25 & 20.769 $\pm$ 0.023 & 1.433 $\pm$ 0.028\\
26 & 20.897 $\pm$ 0.024 & 0.562 $\pm$ 0.040\\
27 & 21.069 $\pm$ 0.025 & 0.600 $\pm$ 0.047\\
28 & 15.345 $\pm$ 0.007 & 0.450 $\pm$ 0.010\\
29 & 20.812 $\pm$ 0.022 & 1.486 $\pm$ 0.027\\
30 & 22.011 $\pm$ 0.057 & 1.343 $\pm$ 0.068\\
31 & 18.395 $\pm$ 0.007 & 0.700 $\pm$ 0.012\\
32 & 19.860 $\pm$ 0.012 & 1.044 $\pm$ 0.017\\
33 & 21.143 $\pm$ 0.042 & 0.973 $\pm$ 0.065\\
34 & 18.571 $\pm$ 0.008 & 1.387 $\pm$ 0.011\\
35 & 21.478 $\pm$ 0.037 & 1.060 $\pm$ 0.051\\
36 & 19.707 $\pm$ 0.011 & 0.978 $\pm$ 0.015\\
37 & 18.523 $\pm$ 0.008 & 0.342 $\pm$ 0.011\\
38 & 18.814 $\pm$ 0.007 & 0.317 $\pm$ 0.011\\
39 & 17.147 $\pm$ 0.007 & 0.464 $\pm$ 0.010\\
\enddata
\end{deluxetable}

\begin{deluxetable}{rccc}
\tablewidth{0pt}   
\tablecaption{ Photometric Sequence Near SN~1999S \label{ta:std_sn99S}}
\tablehead{
\colhead{ID} &
\colhead{$R$}  &
\colhead{$R-I$}} \startdata
1 & 20.469 $\pm$ 0.031 & 0.527 $\pm$ 0.025\\
2 & 20.280 $\pm$ 0.018 & 1.502 $\pm$ 0.015\\
3 & 20.757 $\pm$ 0.027 & 1.397 $\pm$ 0.024\\
4 & 19.973 $\pm$ 0.016 & 0.949 $\pm$ 0.014\\
5 & 16.605 $\pm$ 0.002 & 0.949 $\pm$ 0.002\\
6 & 19.440 $\pm$ 0.014 & 0.485 $\pm$ 0.011\\
7 & 20.148 $\pm$ 0.017 & 1.180 $\pm$ 0.015\\
8 & 19.202 $\pm$ 0.012 & 0.462 $\pm$ 0.010\\
9 & 15.178 $\pm$ 0.001 & 0.361 $\pm$ 0.001\\
10 & 19.809 $\pm$ 0.012 & 1.479 $\pm$ 0.011\\
11 & 19.587 $\pm$ 0.014 & 0.790 $\pm$ 0.011\\
12 & 15.855 $\pm$ 0.001 & 0.402 $\pm$ 0.001\\
13 & 17.087 $\pm$ 0.004 & 0.320 $\pm$ 0.003\\
14 & 16.212 $\pm$ 0.002 & 0.331 $\pm$ 0.001\\
15 & 19.475 $\pm$ 0.016 & 0.370 $\pm$ 0.012\\
16 & 18.980 $\pm$ 0.009 & 0.965 $\pm$ 0.007\\
17 & 16.439 $\pm$ 0.002 & 0.453 $\pm$ 0.001\\
\enddata
\end{deluxetable}

\begin{deluxetable}{rccc}
\tablewidth{0pt}   
\tablecaption{Photometric Sequence Near SN~1999U \label{ta:std_sn99U}}
\tablehead{
\colhead{ID} &
\colhead{$R$}  &
\colhead{$R-I$}} \startdata
1 & 20.771 $\pm$ 0.038 & 1.058 $\pm$ 0.030\\
2 & 18.338 $\pm$ 0.004 & 1.496 $\pm$ 0.004\\
3 & 19.614 $\pm$ 0.023 & 0.347 $\pm$ 0.017\\
4 & 21.069 $\pm$ 0.036 & 1.787 $\pm$ 0.032\\
5 & 19.997 $\pm$ 0.032 & 0.280 $\pm$ 0.024\\
6 & 19.716 $\pm$ 0.011 & 1.673 $\pm$ 0.010\\
7 & 18.893 $\pm$ 0.006 & 1.578 $\pm$ 0.006\\
8 & 15.808 $\pm$ 0.001 & 0.399 $\pm$ 0.001\\
9 & 20.916 $\pm$ 0.036 & 1.341 $\pm$ 0.030\\
10 & 18.763 $\pm$ 0.009 & 1.159 $\pm$ 0.008\\
11 & 19.397 $\pm$ 0.011 & 0.890 $\pm$ 0.009\\
12 & 19.376 $\pm$ 0.018 & 0.345 $\pm$ 0.013\\
13 & 21.127 $\pm$ 0.064 & 0.572 $\pm$ 0.049\\
14 & 19.830 $\pm$ 0.013 & 1.611 $\pm$ 0.011\\
15 & 19.156 $\pm$ 0.012 & 0.778 $\pm$ 0.009\\
16 & 21.241 $\pm$ 0.046 & 1.388 $\pm$ 0.038\\
17 & 21.068 $\pm$ 0.052 & 0.909 $\pm$ 0.042\\
18 & 19.592 $\pm$ 0.025 & 0.373 $\pm$ 0.020\\
19 & 21.564 $\pm$ 0.062 & 1.253 $\pm$ 0.051\\
20 & 21.057 $\pm$ 0.042 & 1.280 $\pm$ 0.034\\
21 & 20.042 $\pm$ 0.017 & 1.286 $\pm$ 0.014\\
22 & 20.927 $\pm$ 0.044 & 1.056 $\pm$ 0.036\\
23 & 21.405 $\pm$ 0.049 & 1.540 $\pm$ 0.041\\
24 & 21.816 $\pm$ 0.084 & 1.259 $\pm$ 0.070\\
25 & 20.807 $\pm$ 0.061 & 0.314 $\pm$ 0.046\\
26 & 20.616 $\pm$ 0.045 & 0.504 $\pm$ 0.035\\
27 & 19.992 $\pm$ 0.033 & 0.504 $\pm$ 0.026\\
28 & 17.728 $\pm$ 0.004 & 1.294 $\pm$ 0.003\\
29 & 20.529 $\pm$ 0.022 & 1.611 $\pm$ 0.020\\
30 & 18.014 $\pm$ 0.005 & 0.516 $\pm$ 0.004\\
31 & 19.972 $\pm$ 0.031 & 0.248 $\pm$ 0.023\\
32 & 21.121 $\pm$ 0.085 & 0.293 $\pm$ 0.063\\
33 & 17.912 $\pm$ 0.005 & 0.958 $\pm$ 0.004\\
34 & 18.811 $\pm$ 0.012 & 1.044 $\pm$ 0.010\\
\enddata
\end{deluxetable}

\begin{deluxetable}{lrrrccc}
\tablewidth{0pt}
\tablecaption{Template Images\label{ta:templates}}
\tablehead{
\colhead{SN name} &
\colhead{Passband} &
\colhead{Telescope/Instrument} &
\colhead{Exposure\tablenotemark{a}} &
\colhead{UT Date} &
\colhead{Seeing\tablenotemark{b}}
}
\startdata
1999M    & $R_c $ & CTIO 4.0m/BTC                     &   900 & 1997/03/09            & 0.98  & \\
         & $I_c $ & VLT-1/FORS-1                      &  1800 & 2000/02/05            & 0.86  & \\
\tableline
1999N    & $R_c $ & CTIO 4.0m/BTC                     &   900 & 1997/03/09            & 1.00  & \\
         & $I_c $ & VLT-1/FORS-1                      &  1800 & 2000/02/07            & 0.84  & \\
\tableline
1999Q    & $F650W$ & HST/WFPC2                         &  1600 & 2000/03/14            & 0.15  & \\
         & $F650W$ & HST/WFPC2 (v.2)\tablenotemark{c}  & 11200 & 1999/02/07-2000/03/14 & 0.55  & \\
         & $F814W$ & HST/WFPC2                         &  2400 & 2000/03/14            & 0.15  & \\
         & $F814W$ & HST/WFPC2 (v.2)\tablenotemark{c}  & 16800 & 1999/02/07-2000/03/14 & 0.55  & \\
\tableline
1999S    & $R_c $ & UH 2.2m/CCD Camera                &  1080 & 1999/12/18            & 0.83  & \\
         & $R_c $ & ESO 2.2m/WFI                      &   900 & 2001/04/20            & 0.90  & \\
         & $R_c $ & CTIO 4.0m/Mosaic II               &   720 & 2001/04/13            & 0.97  & \\
         & $I_c $ & CTIO 4.0m/BTC                     &   780 & 1998/01/24            & 0.95  & \\
\tableline
1999U    & $F650W$ & HST/WFPC2                         &  1600 & 2000/03/14            & 0.15  & \\
         & $F650W$ & HST/WFPC2 (v.2)\tablenotemark{c}  & 11200 & 1999/02/07-2000/03/14 & 0.55  & \\
         & $R_c $ & Baade/LDSS2                       &   360 & 2002/01/18            & 0.90  & \\
         & $I_c $ & CTIO 4.0m/BTC                     &   780 & 1998/01/23            & 1.12  & \\
         & $F814W$ & HST/WFPC2                         &  2400 & 2000/03/14            & 0.15   & \\
         & $F814W$ & HST/WFPC2 (v.2)\tablenotemark{c}  & 16800 & 1999/02/07-2000/03/14 & 0.55  & \\
\enddata
\tablenotetext{a} {Total exposure time in seconds.}
\tablenotetext{b} {FWHM of PSF in arcseconds.}
\tablenotetext{c} {Template image version 2 built from all {\it HST} observations (see text).}
\end{deluxetable}

\begin{deluxetable}{lcrrll}
\tablewidth{0pt}   
\tablecaption{Photometry and $K$-corrections of SN 1999M \label{ta:data99M}}
\tablehead{
\colhead{Date\tablenotemark{a}} &
\colhead{$m_{obs}$\tablenotemark{b}} &
\colhead{K$_{ro}$\tablenotemark{c}} &
\colhead{Exp. (s)\tablenotemark{d}} &
\colhead{Telescope} &
\colhead{Filter}}
\startdata
          &              & K$_{Passband\rightarrow B}$ & & & \\
191.85 & 23.230 $\pm$ 0.050 & -0.717 $\pm$  0.008 &  450 & CTIO 4.0m & $R_c$ \\
196.81 & 23.125 $\pm$ 0.059 & -0.710 $\pm$  0.008 &  300 & CTIO 4.0m & $R_c$ \\
198.78 & 23.251 $\pm$ 0.090 & -0.719 $\pm$  0.009 & 2440 & CTIO 1.5m & B45 \\
225.98 & 24.178 $\pm$ 0.068 & -0.719 $\pm$  0.008 & 1800 & UH 2.2m & $R_c$ \\
          &              & K$_{Passband\rightarrow V}$ & & & \\
198.79 & 22.100 $\pm$ 0.153 & -0.843 $\pm$  0.013 & 1840 & CTIO 1.5m & V45 \\
225.99 & 23.165 $\pm$ 0.105 & -0.840 $\pm$  0.008 & 3600 & UH 2.2m & $I_c$ \\
\enddata
\tablenotetext{a}{MJD $-$ 2451000.}
\tablenotetext{b}{Observed-frame magnitudes.}
\tablenotetext{c}{K-correction computed with PRES light-curve fitting.}
\tablenotetext{d}{Cumulative exposure time in seconds.}

\end{deluxetable}

\begin{deluxetable}{lcrrrll}
\tablewidth{0pt}   
\tablecaption{Photometry and K-corrections of SN 1999N \label{ta:data99N}}
\tablehead{
\colhead{Date\tablenotemark{a}} &
\colhead{$m_{obs}$\tablenotemark{b}} &
\colhead{K$_{ro}$\tablenotemark{c}} &
\colhead{K$_{ro}$\tablenotemark{d}} &
\colhead{Exp. (s)\tablenotemark{e}} &
\colhead{Telescope} &
\colhead{Filter}}
\startdata
          &              & K$_{Passband\rightarrow B}$ & K$_{Passband\rightarrow B}$ & & \\
191.85 & 23.150 $\pm$ 0.028 & -0.674 $\pm$ 0.029 & -0.690 $\pm$ 0.005 &  450 & CTIO 4.0m & $R_c$ \\
196.81 & 23.139 $\pm$ 0.046 & -0.664 $\pm$ 0.029 & -0.692 $\pm$ 0.013 &  300 & CTIO 4.0m & $R_c$ \\
198.81 & 23.268 $\pm$ 0.122 & -0.678 $\pm$ 0.031 & -0.686 $\pm$ 0.006 & 1200 & CTIO 1.5m & B45 \\
219.06 & 24.755 $\pm$ 0.068 & -0.636 $\pm$ 0.029 & -0.649 $\pm$ 0.013 &  300 & Keck-2 & $R$\tablenotemark{f} \\
225.01 & 25.006 $\pm$ 0.125 & -0.633 $\pm$ 0.029 & -0.665 $\pm$ 0.001 & 3000 & UH 2.2m & $R_c$ \\
          &              & K$_{Passband\rightarrow V}$ & K$_{Passband\rightarrow V}$ & & \\
207.86 & 23.458 $\pm$ 0.502 & -0.830 $\pm$ 0.031 & -0.853 $\pm$ 0.005 & 1080 & CTIO 1.5m & V45 \\
219.06 & 24.544 $\pm$ 0.265 & -0.806 $\pm$ 0.020 & -0.825 $\pm$ 0.009 &  300 & Keck-2 & $I$\tablenotemark{f} \\
225.01 & 24.272 $\pm$ 0.434 & -0.781 $\pm$ 0.020 & -0.825 $\pm$ 0.018 & 3600 & UH 2.2m & $I_c$ \\
\enddata
\tablenotetext{a}{MJD $-$ 2451000.}
\tablenotetext{b}{Observed-frame magnitudes.}
\tablenotetext{c}{K-correction computed with PRES light-curve fitting.}
\tablenotetext{d}{K-correction computed with MLCS2k2 light-curve fitting.}
\tablenotetext{e}{Cumulative exposure time in seconds.}
\tablenotetext{f}{Keck-2 passbands have non-standard transmission curves.}

\end{deluxetable}

\begin{deluxetable}{lcrrrll}
\tablewidth{0pt}   
\tablecaption{Photometry and K-corrections of SN 1999Q \label{ta:data99Q}}
\tablehead{
\colhead{Date\tablenotemark{a}} &
\colhead{$m_{obs}$\tablenotemark{b}} &
\colhead{K$_{ro}$\tablenotemark{c}} &
\colhead{K$_{ro}$\tablenotemark{d}} &
\colhead{Exp. (s)\tablenotemark{e}} &
\colhead{Telescope} &
\colhead{Filter}}
\startdata
          &              & K$_{Passband\rightarrow B}$ & K$_{Passband\rightarrow B}$ & & \\
197.690 & 22.417 $\pm$ 0.017 & -0.712 $\pm$ 0.008 & -0.702 $\pm$ 0.001 & 2700 & APO 3.5m & B45 \\
198.630 & 22.479 $\pm$ 0.034 & -0.713 $\pm$ 0.008 & -0.702 $\pm$ 0.001 & 4520 & CTIO 1.5m & B45 \\
199.740 & 22.573 $\pm$ 0.018 & -0.715 $\pm$ 0.008 & -0.702 $\pm$ 0.001 & 2640 & ESO 3.6m & B45 \\
210.223 & 23.048 $\pm$ 0.037 & -0.844 $\pm$ 0.040 & -0.743 $\pm$ 0.026 & 1600 & HST & F675W \\
216.219 & 23.405 $\pm$ 0.047 & -0.891 $\pm$ 0.040 & -0.819 $\pm$ 0.040 & 1600 & HST & F675W \\
219.000 & 23.807 $\pm$ 0.040 & -0.787 $\pm$ 0.010 & -0.714 $\pm$ 0.001 &  500 & Keck-2 & $R$\tablenotemark{f} \\
219.610 & 23.806 $\pm$ 0.048 & -0.733 $\pm$ 0.008 & -0.710 $\pm$ 0.001 & 3600 & APO 3.5m & B45 \\
223.449 & 23.872 $\pm$ 0.066 & -0.934 $\pm$ 0.040 & -0.912 $\pm$ 0.013 & 1600 & HST & F675W \\
227.760 & 24.160 $\pm$ 0.054 & -0.807 $\pm$ 0.010 & -0.717 $\pm$ 0.001 &  720 & Keck-2 & $R$\tablenotemark{f} \\
230.730 & 24.253 $\pm$ 0.089 & -0.956 $\pm$ 0.040 & -0.940 $\pm$ 0.005 & 1600 & HST & F675W \\
237.430 & 24.569 $\pm$ 0.115 & -0.955 $\pm$ 0.040 & -0.930 $\pm$ 0.001 & 1600 & HST & F675W \\
244.422 & 24.817 $\pm$ 0.146 & -0.941 $\pm$ 0.040 & -0.947 $\pm$ 0.005 & 1600 & HST & F675W \\
          &              & K$_{Passband\rightarrow V}$ & K$_{Passband\rightarrow V}$ & & \\
192.690 & 22.279 $\pm$ 0.027 & -0.857 $\pm$ 0.008 & -0.863 $\pm$ 0.001 & 2340 & CTIO 4.0m & $I_c$ \\
199.670 & 22.290 $\pm$ 0.037 & -0.836 $\pm$ 0.008 & -0.838 $\pm$ 0.003 & 4320 & ESO 3.6m & V45 \\
210.289 & 22.668 $\pm$ 0.030 & -0.771 $\pm$ 0.013 & -0.791 $\pm$ 0.006 & 2400 & HST & F814W \\
216.270 & 22.871 $\pm$ 0.032 & -0.773 $\pm$ 0.013 & -0.788 $\pm$ 0.008 & 2400 & HST & F814W \\
219.010 & 23.010 $\pm$ 0.072 & -0.853 $\pm$ 0.008 & -0.825 $\pm$ 0.014 &  360 & Keck-2 & $I$\tablenotemark{f} \\
219.650 & 22.945 $\pm$ 0.080 & -0.829 $\pm$ 0.008 & -0.869 $\pm$ 0.004 & 1800 & ESO 3.6m & $I_c$ \\
223.473 & 23.149 $\pm$ 0.040 & -0.779 $\pm$ 0.013 & -0.804 $\pm$ 0.007 & 2400 & HST & F814W \\
224.850 & 23.217 $\pm$ 0.090 & -0.861 $\pm$ 0.008 & -0.875 $\pm$ 0.003 & 2400 & UH 2.2m & $I_c$ \\
225.810 & 23.245 $\pm$ 0.110 & -0.863 $\pm$ 0.008 & -0.876 $\pm$ 0.003 & 2700 & UH 2.2m & $_c$ \\
230.784 & 23.374 $\pm$ 0.048 & -0.778 $\pm$ 0.013 & -0.820 $\pm$ 0.007 & 2400 & HST & F814W \\
237.445 & 23.675 $\pm$ 0.058 & -0.769 $\pm$ 0.013 & -0.843 $\pm$ 0.003 & 2400 & HST & F814W \\
244.432 & 23.959 $\pm$ 0.071 & -0.760 $\pm$ 0.013 & -0.826 $\pm$ 0.007 & 2400 & HST & F814W \\
\enddata
\tablenotetext{a}{MJD $-$ 2451000.}
\tablenotetext{b}{Observed-frame magnitudes.}
\tablenotetext{c}{K-correction computed with PRES light-curve fitting.}
\tablenotetext{d}{K-correction computed with MLCS2k2 light-curve fitting.}
\tablenotetext{d}{Cumulative exposure time in seconds.}
\tablenotetext{f}{Keck-2 passbands have non-standard transmission curves.}

\end{deluxetable}

\begin{deluxetable}{lcrrrll}
\tablewidth{0pt}   
\tablecaption{Photometry and K-corrections of SN 1999S \label{ta:data99S}}
\tablehead{
\colhead{Date\tablenotemark{a}} &
\colhead{$m_{obs}$\tablenotemark{b}} &
\colhead{K$_{ro}$\tablenotemark{c}} &
\colhead{K$_{ro}$\tablenotemark{d}} &
\colhead{Exp. (s)\tablenotemark{e}} &
\colhead{Telescope} &
\colhead{Filter}}
\startdata
          &              & K$_{Passband\rightarrow B}$ & K$_{Passband\rightarrow B}$ & & \\
198.700 & 22.563 $\pm$ 0.028 & -0.718 $\pm$  0.005 &  -0.715 $\pm$  0.001 & 3608 & CTIO 1.5m & B45 \\
206.820 & 22.122 $\pm$ 0.016 & -0.717 $\pm$  0.005 &  -0.715 $\pm$  0.001 & 3284 & CTIO 1.5m & B45 \\
219.040 & 22.281 $\pm$ 0.014 & -0.716 $\pm$  0.006 &  -0.721 $\pm$  0.001 &  240 & Keck-2 & $R$\tablenotemark{f} \\
224.750 & 22.509 $\pm$ 0.028 & -0.715 $\pm$  0.005 &  -0.709 $\pm$  0.003 & 2400 & APO 3.5m & B45 \\
225.880 & 22.650 $\pm$ 0.030 & -0.730 $\pm$  0.006 &  -0.728 $\pm$  0.007 & 1800 & UH 2.2m & $R_c$ \\
          &              & K$_{Passband\rightarrow V}$ & K$_{Passband\rightarrow V}$ & & \\
192.770 & 23.421 $\pm$ 0.033 & -0.855 $\pm$  0.005 &  -0.878 $\pm$  0.002 & 2340 & CTIO 4.0m & $I_c$ \\
206.820 & 22.326 $\pm$ 0.061 & -0.849 $\pm$  0.009 &  -0.865 $\pm$  0.006 & 1800 & CTIO 1.5m & V45 \\
219.040 & 22.241 $\pm$ 0.010 & -0.863 $\pm$  0.020  &  -0.825 $\pm$ 0.009 &  450 & Keck-2 & $I$\tablenotemark{f} \\
225.910 & 22.428 $\pm$ 0.016 & -0.852 $\pm$  0.005 &  -0.866 $\pm$  0.001 & 3600 & UH 2.2m & $I_c$ \\
\enddata
\tablenotetext{a}{MJD $-$ 2451000.}
\tablenotetext{b}{Observed-frame magnitudes.}
\tablenotetext{c}{K-correction computed with PRES light-curve fitting.}
\tablenotetext{d}{K-correction computed with MLCS2k2 light-curve fitting.}
\tablenotetext{d}{Cumulative exposure time in seconds.}
\tablenotetext{f}{Keck-2 passbands have non standard transmission curves.}

\end{deluxetable}

\begin{deluxetable}{lcrrrll}
\tablewidth{0pt}   
\tablecaption{Photometry and K-corrections of SN 1999U \label{ta:data99U}}
\tablehead{
\colhead{Date\tablenotemark{a}} &
\colhead{$m_{obs}$\tablenotemark{b}} &
\colhead{K$_{ro}$\tablenotemark{c}} &
\colhead{K$_{ro}$\tablenotemark{d}} &
\colhead{Exp. (s)\tablenotemark{e}} &
\colhead{Telescope} &
\colhead{Filter}}
\startdata
          &              & K$_{Passband\rightarrow B}$ & K$_{Passband\rightarrow B}$ & & \\
197.77 & 22.783 $\pm$ 0.035 & -0.730 $\pm$ 0.020 & -0.716 $\pm$ 0.005 & 1800 & APO 3.5m & B45 \\
198.73 & 22.736 $\pm$ 0.050 & -0.728 $\pm$ 0.020 & -0.714 $\pm$ 0.009 & 2990 & CTIO 1.5m & B45 \\
199.83 & 22.654 $\pm$ 0.027 & -0.727 $\pm$ 0.020 & -0.711 $\pm$ 0.009 & 1980 & ESO 3.6m & B45 \\
210.36 & 22.420 $\pm$ 0.023 & -0.806 $\pm$ 0.019 & -0.811 $\pm$ 0.004 & 1600 & HST & F675W \\
216.41 & 22.650 $\pm$ 0.026 & -0.818 $\pm$ 0.019 & -0.804 $\pm$ 0.001 & 1600 & HST & F675W \\
219.02 & 22.871 $\pm$ 0.021 & -0.681 $\pm$ 0.008 & -0.689 $\pm$ 0.005 &  486 & Keck-1 & $R$\tablenotemark{f} \\
223.53 & 23.050 $\pm$ 0.100 & -0.830 $\pm$ 0.019 & -0.808 $\pm$ 0.005 & 1600 & HST & F675W \\
224.90 & 23.179 $\pm$ 0.053 & -0.680 $\pm$ 0.008 & -0.704 $\pm$ 0.001 & 1800 & UH 2.2m & $R_c$ \\
230.87 & 23.470 $\pm$ 0.044 & -0.838 $\pm$ 0.019 & -0.826 $\pm$ 0.009 & 1600 & HST & F675W \\
237.51 & 23.970 $\pm$ 0.054 & -0.840 $\pm$ 0.019 & -0.839 $\pm$ 0.003 & 1600 & HST & F675W \\
244.49 & 24.440 $\pm$ 0.074 & -0.840 $\pm$ 0.019 & -0.836 $\pm$ 0.003 & 1600 & HST & F675W \\
          &              & K$_{Passband\rightarrow V}$ & K$_{Passband\rightarrow V}$ & & \\
192.73 & 23.038 $\pm$ 0.038 & -0.885 $\pm$ 0.013 & -0.909 $\pm$ 0.004 & 2340 & CTIO 4.0m & $I_c$ \\
199.80 & 22.535 $\pm$ 0.051 & -0.857 $\pm$ 0.021 & -0.876 $\pm$ 0.007 & 3600 & ESO 3.6m & V45 \\
210.38 & 22.320 $\pm$ 0.026 & -0.824 $\pm$ 0.021 & -0.788 $\pm$ 0.001 & 2400 & HST & F814W \\
216.43 & 22.470 $\pm$ 0.027 & -0.804 $\pm$ 0.021 & -0.788 $\pm$ 0.001 & 2400 & HST & F814W \\
219.02 & 22.702 $\pm$ 0.035 & -0.852 $\pm$ 0.013 & -0.845 $\pm$ 0.007 &  480 & Keck-1 & $I$\tablenotemark{f} \\
219.75 & 22.595 $\pm$ 0.055 & -0.831 $\pm$ 0.021 & -0.822 $\pm$ 0.007 & 2400 & APO 3.5m & V45 \\
223.55 & 22.660 $\pm$ 0.062 & -0.771 $\pm$ 0.021 & -0.788 $\pm$ 0.001 & 2400 & HST & F814W \\
224.93 & 22.704 $\pm$ 0.074 & -0.840 $\pm$ 0.013 & -0.858 $\pm$ 0.008 & 3000 & UH 2.2m & $I_c$ \\
230.92 & 23.030 $\pm$ 0.033 & -0.737 $\pm$ 0.021 & -0.788 $\pm$ 0.001 & 2400 & HST & F814W \\
237.52 & 23.380 $\pm$ 0.041 & -0.713 $\pm$ 0.021 & -0.788 $\pm$ 0.001 & 2400 & HST & F814W \\
244.51 & 23.610 $\pm$ 0.047 & -0.693 $\pm$ 0.021 & -0.788 $\pm$ 0.001 & 2400 & HST & F814W \\
\enddata
\tablenotetext{a}{MJD $-$ 2451000.}
\tablenotetext{b}{Observed-frame magnitudes.}
\tablenotetext{c}{K-correction computed with PRES light-curve fitting.}
\tablenotetext{d}{K-correction computed with MLCS2k2 light-curve fitting.}
\tablenotetext{d}{Cumulative exposure time in seconds.}
\tablenotetext{f}{Keck-2 passbands have non standard transmission curves.}

\end{deluxetable}

\begin{deluxetable}{l c c c c c c}
\tablewidth{0pt}   
\tablecaption{Parameters of the Light-Curve Fits (PRES method)\label{ta:table_pres}}
\tablehead{
\colhead{SN} &
\colhead{ $z$} &
\colhead{$E(B-V)_{\rm Gal.}$} &
\colhead{ $\mu$ (mag)} &
\colhead{ $\Delta m_{15}$ (mag)\tablenotemark{a}} &
\colhead{$E(B-V)_{\rm Host}$} &
\colhead{ $t_{max}^{B}$\tablenotemark{b}}
} \startdata
1999M & 0.493 &  0.042  & 41.02 $\pm$ 0.81 &  0.83  $\pm$  0.18  & 0.61 $\pm$ 0.20 & 199.8 \\
1999N & 0.537 &  0.050  & 43.11 $\pm$ 0.20 &  1.22 $\pm$  0.15  & 0.00 $\pm$ 0.03 & 192.2 \\
1999Q & 0.459 &  0.021  & 42.63 $\pm$ 0.14 &  0.83 $\pm$  0.02 &  0.05 $\pm$ 0.03 & 189.6 \\
1999S & 0.474 &  0.033  & 42.55 $\pm$ 0.20 &  0.87 $\pm$  0.02 &  0.00 $\pm$ 0.03 & 210.0 \\
1999U & 0.511 &  0.039  & 42.48 $\pm$ 0.16 &  1.02  $\pm$  0.04  & 0.06 $\pm$ 0.03 & 207.1 \\ \hline
\enddata
\tablenotetext{a}{Magnitudes below rest-frame $B$ maximum.}
\tablenotetext{b}{ JD$_{max}-2,451,000$ days.}
\end{deluxetable}

\begin{deluxetable}{l c c c c c}
\tablewidth{0pt}   
\tablecaption{Parameters of the Light-Curve Fits (MLCS2k2 method)\label{ta:table_mlcs2k2}}
\tablehead{
\colhead{SN} &
\colhead{ $z$} &
\colhead{$E(B-V)_{\rm Gal.}$} &
\colhead{ $\mu$ (mag)} &
\colhead{$E(B-V)_{\rm Host}$} &
\colhead{$t_{max}^{B}$\tablenotemark{a}}
} \startdata
1999N & 0.537 &  0.050  & 42.81 $\pm$ 0.41 &  0.18 & 192.4 \\
1999Q & 0.459 &  0.021  & 42.62 $\pm$ 0.21 &  0.19 & 192.3 \\
1999S & 0.474 &  0.033  & 42.77 $\pm$ 0.21 & -0.07 & 209.6 \\
1999U & 0.511 &  0.039  & 42.80 $\pm$ 0.20 &  0.18 & 205.2 \\ \hline
\enddata
\tablenotetext{a}{ JD$_{max}-2,451,000$ days.}
\end{deluxetable}

\begin{deluxetable}{lcrrrr}
\tablewidth{0pt}
\tablecaption{Fits to $\Lambda$CDM Cosmology\label{ta:cosmology}}
\tablehead{
\colhead{Method} &
\colhead{Constraint} &
\colhead{$\Omega_M$\tablenotemark{a}} &
\colhead{$\Omega_\Lambda$\tablenotemark{a}} &
\colhead{$\chi^2$} &
\colhead{N\tablenotemark{b}}
}
\startdata
\multicolumn{6}{c}{\underline{All HZT SNe except this work}}\\
\vspace{0.1cm}
MLCS2k2    & none &  $0.68^{+0.18}_{-0.26}$ &  $1.20^{+0.34}_{-0.41}$ & 109 & 118 \\
\vspace{0.1cm}
PRES & none &  $0.27^{+0.17}_{-0.19}$ &  $0.88^{+0.29}_{-0.27}$ & 112 & 101 \\
\vspace{0.1cm}
MLCS2k2 & $\Omega_M$ + $\Omega_\Lambda = 1$ &  $0.33^{+0.07}_{-0.05}$ & $1-\Omega_M$ & 107 & 118 \\
\vspace{0.1cm}
PRES    & $\Omega_M$ + $\Omega_\Lambda = 1$ &  $0.20^{+0.05}_{-0.04}$ & $1-\Omega_M$ & 112 & 101 \\
\tableline
\multicolumn{6}{c}{\underline{All HZT SNe}}\\
\vspace{0.1cm}
MLCS2k2 & none &  $0.79^{+0.15}_{-0.18}$ &  $1.57^{+0.24}_{-0.32}$ & 118 & 122 \\
\vspace{0.1cm}
PRES    & none &  $0.43^{+0.17}_{-0.19}$ &  $1.18^{+0.27}_{-0.28}$ & 133 & 106 \\
\vspace{0.1cm}
MLCS2k2 & $\Omega_M$ + $\Omega_\Lambda = 1$ &  $0.29^{+0.06}_{-0.05}$ & $1-\Omega_M$ & 124 & 122 \\
\vspace{0.1cm}
PRES    & $\Omega_M$ + $\Omega_\Lambda = 1$ &  $0.18^{+0.05}_{-0.04}$ & $1-\Omega_M$ & 132 & 106 \\
\tableline
\multicolumn{6}{c}{\underline{SNe of this work plus gold \& silver of Riess et al. (2004)}} \\
\vspace{0.1cm}
MLCS2k2 & none &  $0.50^{+0.10}_{-0.10}$ &  $1.10^{+0.16}_{-0.17}$ & 219 & 187 \\
MLCS2k2 & $\Omega_M$ + $\Omega_\Lambda = 1$ &  $0.29^{+0.04}_{-0.03}$ & $1-\Omega_M$ & 224 & 187 \\
\tableline
\multicolumn{6}{c}{\underline{SNe of this work plus gold of Riess et al. (2004)}} \\
\vspace{0.1cm}
MLCS2k2 & none &  $0.48^{+0.09}_{-0.12}$ &  $1.04^{+0.17}_{-0.19}$ & 184 & 158 \\
MLCS2k2 & $\Omega_M$ + $\Omega_\Lambda = 1$ &  $0.30^{+0.04}_{-0.04}$ & $1-\Omega_M$ & 186 & 158 \\
\enddata
\tablenotetext{a}{$1\sigma$ uncertainties}
\tablenotetext{b}{Number of SNe for this sample and method.}
\end{deluxetable}

\end{document}